\documentclass[oldversion,longauth]{aa}
\def\simlt{\lower.5ex\hbox{$\; \buildrel < \over \sim \;$}}
\def\simgt{\lower.5ex\hbox{$\; \buildrel > \over \sim \;$}}
\usepackage{graphicx}
\begin{document}
\title{The VIMOS VLT Deep Survey
         \thanks{based on data
         obtained with the European Southern Observatory Very Large
         Telescope, Paranal, Chile, program 070.A-9007(A), and on data
         obtained at the Canada-France-Hawaii Telescope, operated by
         the CNRS of France, CNRC in Canada and the University of Hawaii}
}
\subtitle{The Assembly History of the Stellar Mass in Galaxies:
from the Young to the Old Universe 
}

\author{
%ANALYSIS
     L. Pozzetti    \inst{1} 
\and M. Bolzonella  \inst{1} 
\and F. Lamareille \inst{1,9,6}
\and G. Zamorani \inst{1} 
\and P. Franzetti \inst{2}
\and O. Le F\`evre \inst{3}
\and A. Iovino \inst{4}
\and S. Temporin \inst{4}
\and O. Ilbert \inst{5}
\and S. Arnouts \inst{3}
\and S. Charlot \inst{6,7}
\and J. Brinchmann  \inst{8} 
\and E. Zucca    \inst{1}
\and L. Tresse \inst{3}
\and M. Scodeggio \inst{2}
\and L. Guzzo \inst{4}
%BUILDERS(b):
\and D. Bottini \inst{2}
\and B. Garilli \inst{2}
\and V. Le Brun \inst{3}
\and D. Maccagni \inst{2}
\and J.P. Picat \inst{9}
\and R. Scaramella \inst{10,11}
\and G. Vettolani \inst{10}
\and A. Zanichelli \inst{10}
%SURVEY CORE(s):
\and C. Adami \inst{3}
\and S. Bardelli    \inst{1}
\and A. Cappi    \inst{1}
\and P. Ciliegi    \inst{1}  
\and T. Contini \inst{9}
\and S. Foucaud \inst{12}
\and I. Gavignaud \inst{13}
\and H.J. McCracken \inst{7,14}
\and B. Marano     \inst{15}  
\and C. Marinoni \inst{16}
\and A. Mazure \inst{3}
\and B. Meneux \inst{2,4}
\and R. Merighi   \inst{1} 
\and S. Paltani \inst{17,18}
\and R. Pell\`o \inst{9}
\and A. Pollo \inst{3,19}
\and M. Radovich \inst{20}
%ASSOCIATES(a):
\and M. Bondi \inst{10}
\and A. Bongiorno \inst{15}
\and O. Cucciati \inst{4,21}
\and S. de la Torre \inst{3}
\and L. Gregorini \inst{22,10}
\and Y. Mellier \inst{7,14}
\and P. Merluzzi \inst{20}
\and D. Vergani \inst{2}
\and C.J. Walcher \inst{3}
}
   
   \offprints{ Lucia Pozzetti \email{lucia.pozzetti@oabo.inaf.it} }

\institute{
%1) 
INAF-Osservatorio Astronomico di Bologna - Via Ranzani,1, I-40127, Bologna, Italy
\and
%2)
INAF-IASF - via Bassini 15, I-20133, Milano, Italy
\and
%3)
Laboratoire d'Astrophysique de Marseille, UMR 6110 CNRS-Universit\'e de
Provence,  BP8, 13376 Marseille Cedex 12, France
\and
%4)
INAF-Osservatorio Astronomico di Brera - Via Brera 28, Milan, Italy
\and
%5)
Institute for Astronomy, 2680 Woodlawn Dr., University of Hawaii,
Honolulu, Hawaii, 96822
\and	
%6)
 Max Planck Institut fur Astrophysik, 85741, Garching, Germany
\and
%7)
 Institut d'Astrophysique de Paris, UMR 7095, 98 bis Bvd Arago, 75014
Paris, France
\and
%8)
 Centro de Astrofísica da Universidade do Porto, Rua das Estrelas,
4150-762 Porto, Portugal 
\and
%9)
 Laboratoire d'Astrophysique de Toulouse/Tabres (UMR5572), CNRS, Universit\'e Paul Sabatier -
 Toulouse III, Observatoire Midi-Pyri\'en\'ees, 14 av. E. Belin, F-31400 Toulouse, France 
\and
%10)
INAF-IRA - Via Gobetti,101, I-40129, Bologna, Italy
\and
%11)
INAF-Osservatorio Astronomico di Roma - Via di Frascati 33,
I-00040, Monte Porzio Catone, Italy
\and
%%12)
  School of Physics \& Astronomy, University of Nottingham, University Park, Nottingham, NG72RD, UK
\and
%%13)
 Astrophysical Institute Potsdam, An der Sternwarte 16, D-14482
Potsdam, Germany
\and
%%14)
 Observatoire de Paris, LERMA, 61 Avenue de l'Observatoire, 75014 Paris, 
France
\and
%%15)
 Universit\`a di Bologna, Dipartimento di Astronomia - Via Ranzani,1,
I-40127, Bologna, Italy
\and
%%16)
Centre de Physique Th\'eorique, UMR 6207 CNRS-Universit\'e de Provence, 
F-13288 Marseille France
\and
%17)
Integral Science Data Centre, ch. d'\'Ecogia 16, CH-1290 Versoix
\and
%18)
Geneva Observatory, ch. des Maillettes 51, CH-1290 Sauverny, Switzerland
\and
%19)
Astronomical Observatory of the Jagiellonian University, ul Orla 171, 
30-244 Krak{\'o}w, Poland
\and
%20)
INAF-Osservatorio Astronomico di Capodimonte - Via Moiariello 16, I-80131, Napoli,
Italy
\and
%21)
Universit\'a di Milano-Bicocca, Dipartimento di Fisica - 
Piazza delle Scienze, 3, I-20126 Milano, Italy
\and
%%22)
 Universit\`a di Bologna, Dipartimento di Fisica - Via Irnerio 46,
I-40126, Bologna, Italy
}

\authorrunning {L.Pozzetti et al.}

\titlerunning {The VVDS: The Galaxy Stellar Mass Assembly History}

\date{Received 04 04 2007; accepted 18 08 2007}

\abstract{ 
We present a detailed analysis of the Galaxy Stellar Mass
Function (GSMF) of galaxies up to $z=2.5$ as obtained from the 
VIMOS VLT Deep Survey (VVDS).
Our survey offers
the possibility to investigate it using 
two different samples: (1) an optical ($I$-selected $17.5<I_{\rm AB}<24$) main spectroscopic
sample of about 6500 galaxies over 1750 arcmin$^2$ and 
(2) a near-IR ($K$-selected $K_{\rm AB}<22.34 ~\&~ K_{\rm AB}<22.84$) sample of about 10200 galaxies, 
with photometric redshifts accurately calibrated on the VVDS spectroscopic sample,
over 610 arcmin$^2$.
We apply and compare two
different methods to estimate the stellar mass ${\cal M}_{\rm stars}$ from broad-band
photometry based on different assumptions on the galaxy
star-formation history.
We find that the accuracy of the photometric stellar mass
is overall satisfactory, and show that the addition of secondary bursts to
a continuous star formation history produces systematically higher (up to 40\%) stellar masses.
We derive the cosmic evolution of the GSMF, the galaxy number density and the stellar mass density
in different mass ranges.
At low redshift ($z\simeq0.2$) we find a substantial population of low-mass galaxies ($<10^9 M_\odot$) 
composed by faint blue galaxies ($M_I-M_K \simeq 0.3$).
In general the stellar mass function evolves slowly 
up to $z\sim0.9$ and more significantly above this redshift, in particular for low mass systems.
Conversely, a massive population is present up to $z=2.5$ and have extremely red colours ($M_I-M_K\simeq 0.7-0.8$).
We find a decline with redshift of the overall number density of galaxies for all masses
($59\pm5$\% for ${\cal M}_{\rm stars} > 10^8 M_\odot$ at $z=1$), and a mild mass-dependent average evolution (`mass-downsizing').
In particular our data are consistent with mild/negligible ($<30$\%)  evolution up to $z\sim0.7$
for massive galaxies ($>6\times10^{10} M_\odot$). 
For less massive systems the no-evolution scenario is excluded.
Specifically, a large fraction ($\ge50\%$) of massive galaxies
have been already assembled and converted
most of their gas into stars at $z\sim1$, ruling out
the `dry mergers' as the major mechanism of their assembly history below $z\simeq1$.
This fraction decreases to $\sim33\%$ at $z\sim2$.
Low-mass systems have decreased continuously in number
density (by a factor up to $4.1\pm0.9$) from the present age to $z=2$, 
consistently with a prolonged mass assembly also at $z<1$.
The evolution of the stellar mass density
is relatively slow with redshift, with a decrease of 
a factor $2.3\pm0.1$ at $z=1$ and about $4.5\pm0.3$ at $z=2.5$.

         }

\maketitle

%-------------------------------------------------------------------------------
\section{Introduction}\label{sec:intro}

One of the main and still open question of modern cosmology
is how and when galaxies formed and in particular when they assembled their
stellar mass.  There are growing but still controversial evidences in near-IR (NIR)
surveys that luminous and rather massive old galaxies were 
quite common already at $z\sim1$ (Pozzetti et al. 2003, Fontana et al. 2004, 
Saracco et al. 2004, 2005, Caputi et al. 2006a) and up to $z\sim2$
(Cimatti et al. 2004, Glazebrook et al 2004). These surveys indicate that a significant fraction of
early-type massive galaxies were already in place
at least up to $z\sim1$. Therefore they should have
formed their stars and assembled their stellar mass at higher
redshifts. As in the local universe, at $z\simeq 1.5$ 
these galaxies still dominate the near-IR
luminosity function and stellar mass density of the universe 
(Pozzetti et al. 2003, Fontana et al. 2004, Strazzullo et al. 2006).
These results favour a high-$z$ mass assembly, in particular for massive galaxies, 
in apparent contradiction with the
hierarchical scenario of galaxy formation, applied to both dark and
baryonic matter, which predicts that galaxies form 
through merging at later cosmic time.  In these models massive galaxies,
in particular, assembled most of their stellar mass via merging 
only at $z<1$ (De Lucia et al. 2006).  
From several observations it seems that baryonic
matter has a mass-dependent assembly history, from massive to small objects,
(i.e. the `downsizing' scenario in star formation,
firstly defined by Cowie et al. 1996, is valid also for mass assembly), 
opposite to the dark matter (DM) halos assembly. 
The continuous merging of DM halos in the hierarchical models, indeed, should result in 
an `upsizing' in mass assembly, with the most massive galaxies being the last to be fully 
assembled. 
If we trust the hierarchical $\Lambda$CDM universe, 
the source of this discrepancy between observations and simple basic models could be due
to the difficult physical treatment of the baryonic component, such as the star formation 
history/timescale, feedback, dust content, AGN feedback
or to a missing ingredient 
in the hierarchical models of galaxy formation (for the inclusion of AGN feedback, see Bower et al. 2006, 
Kang et al. 2006, De Lucia \& Blaizot 2007, Menci et al. 2006, Monaco et al. 2006; 
and see Neistein et al. 2006 for the description of
a natural downsizing in star formation in the hierarchical galaxy formation models
and a recent review by Renzini 2007). 

Considering optically selected surveys, a strong number density evolution
of early type galaxies has been recently reported from the COMBO17 and DEEP2 surveys
(Bell et al. 2004, Faber et al. 2005), 
with a corresponding increase by a factor 2 of 
their stellar mass since $z\sim1$, possibly due to so 
called `dry-mergers' 
(even if the observational results on major merging and dry-merging 
are still contradictory, see Bell et al. 2006, van Dokkum 2005, Lin et al. 2004
and Renzini 2007 for a summary).
This is at variance with results from the VIMOS-VLT Deep Survey
(VVDS, Le~F\`evre et al. \cite{lefevre03}), conducted at 
greater depth and using spectroscopic redshifts in a 
large contiguous area. 
From the VVDS, Zucca et al. (2006) found that the $B$-band luminosity 
function of early type galaxies is consistent with passive evolution
up to $z \sim 1.1$, while the number of bright ($M_{B_{AB}} < -20$)
early type galaxies has decreased only by $\sim 40$\% from $z\sim 0.3$
to $z\sim 1.1$. 
Similarly, Brown et al. (2007), in the NOAO Deep Wide Field survey over $\sim 10$ deg$^2$, 
found that the $B$-band luminosity density of $L^*$ galaxies increases by only 
$36\pm13\%$ from $z=0$ to $z=1$ and conclude that mergers do not produce rapid growth of luminous red 
galaxy stellar masses between $z=1$ and the present day.

The VVDS is very well suited for this 
kind of studies, thanks to its depth and wide area, covered by
multi-wavelength photometry and deep spectroscopy.  
The simple $17.5<I_{AB} <24$ VVDS magnitude limit selection is significantly
fainter than other complete spectroscopic surveys and allows
the determination of the faint and low mass population
with unprecedented accuracy. 
Most of the previous existing surveys are instead very small 
and/or not deep enough, or based
only on photometric redshifts.  

Given the still controversial results based on morphology or colour-selected
early-type galaxies (see Franzetti et al. 2007 for a discussion on colour-selected contamination),
we prefer to study the total galaxy population using the stellar mass content.
Here we present results on the cosmic evolution of the Galaxy Stellar
Mass Function (GSMF) and mass density to $z=2.5$ in the deep VVDS
spectroscopic survey, limited to $17.5<I_{\rm AB}<24$, over 
$\sim 1750$ arcmin$^2$ and based on about 6500 galaxies 
with secure spectroscopic 
redshifts and multiband (from UV to near-IR) photometry. In addition,
we derive the GSMF also for a $K$-selected sample based on
about 6600 galaxies ($K_{\rm AB}<22.34$) 
in an area of 442 arcmin$^2$ and 
about 3600 galaxies in a deeper ($K_{\rm AB}<22.84$)  smaller
area of 168 arcmin$^2$, making use of photometric redshifts, accurately calibrated on the VVDS
spectroscopic sample, and spectroscopic redshifts when available.  

Throughout the paper we adopt the cosmology $\Omega_m = 0.3$ and
$\Omega_\Lambda = 0.7$, with $h_{70} = H_0 / 70$ km s$^{-1}$
Mpc$^{-1}$. Magnitudes are given in the AB system and the suffix AB will be dropped
from now on.

%-------------------------------------------------------------------------------
\section{The First Epoch VVDS Sample}\label{sec:sample}

The VVDS is an ongoing program aiming to map the evolution of galaxies,
large scale structures and AGN through redshift measurements 
of $\sim 10^5$ objects, obtained with the VIsible Multi-Object
Spectrograph (VIMOS, Le~F\`evre et al. \cite{vimos}), mounted on the
ESO Very Large Telescope (UT3), in combination with a multi-wavelength
dataset from radio to X-rays. The VVDS is described in detail in Le~F\`evre et
al. (\cite{vvds1}). Here we summarize only the main characteristics of
the survey.

The VVDS is made of a wide part, with spectroscopy in the range
$17.5\le I \le 22.5$ on 4 fields ($\sim 2\times 2$ deg$^2$ each), and a deep part, with spectroscopy in the range
$17.5\le I\le 24$ on the field 0226-04 (F02 hereafter). Multicolour
photometry is available for each field (Le~F\`evre et al.  
\cite{photom1}). In particular, the $B$, $V$, $R$, $I$ photometry
for the 0226-04 deep field, covering $\sim 1$ deg$^2$, has been obtained
at CFHT and is described in detail in McCracken et al. (\cite{photom2}).
The photometric depth reached in this field is 26.5, 26.2, 25.9, 25.0
(50\% completeness for point-like sources), respectively in the $B$, $V$, $R$, $I$ bands. 
Moreover, $U<25.4$ (50\% completeness) photometry obtained with the
WFI at the ESO-2.2m telescope (Radovich et al. \cite{photomU})
and $K_s$ band (hereafter $K$) photometry with NTT+SOFI at the depth
(50\% completeness) of 23.34 (Temporin et al. in preparation) are available for
wide sub-areas of this field. Moreover, an area of about
$170$ arcmin$^2$ has been covered by
deeper $J$ and $K$ band observations with NTT+SOFI at the depth
(50\% completeness) of 24.15 and 23.84, respectively (Iovino et
al. \cite{photomK}). 
The deep F02 field has been observed also by the CFHT Legacy
Survey (CFHTLS\footnote{Based on observations obtained with MegaPrime/MegaCam, a joint  
project of CFHT and CEA/DAPNIA, at the Canada-France-Hawaii Telescope  
(CFHT) which is operated by the National Research Council (NRC) of  
Canada, the Institut National des Science de l'Univers of the Centre  
National de la Recherche Scientifique (CNRS) of France, and the  
University of Hawaii. This work is based in part on data products  
produced at TERAPIX and the Canadian Astronomy Data Centre as part of  
the Canada-France-Hawaii Telescope Legacy Survey, a collaborative  
project of NRC and CNRS.}) in several optical bands ($u^*$, $g'$, $r'$, $i'$,
$z'$) at very faint depth ($u^*=26.4, g'=26.3, r'=26.1, i'=25.9, z'=24.9$,
50\% completeness). 

Spectroscopic observations of a randomly selected subsample of
objects in an area of $\sim 0.5$ deg$^2$, with an average 
sampling rate of about 25\%, were performed in the F02 field with
VIMOS at the VLT. 

Spectroscopic data were reduced with the VIMOS Interactive Pipeline
Graphical Interface (VIPGI, Scodeggio et al. \cite{vipgi}, Zanichelli
et al.  \cite{ifu}) and redshift measurements were performed with 
an automatic package (KBRED)
and then visually checked. Each redshift measurement was assigned a quality flag,
ranging from 0 (failed measurement) to 4 (100\% confidence level);
flag 9 indicates spectra with a single emission line, for which
multiple redshift solutions are possible. Further details on the quality flags
are given in Le~F\`evre et al. (\cite{vvds1}).

The analysis presented in this paper is based on the first epoch VVDS
deep sample, which has been obtained from the first spectroscopic observations (fall
2002) on the field VVDS-02h, which cover
1750 arcmin$^2$.

%-------------------------------------------------------------------------------
\begin{figure}
\centering
\includegraphics[width=\hsize]{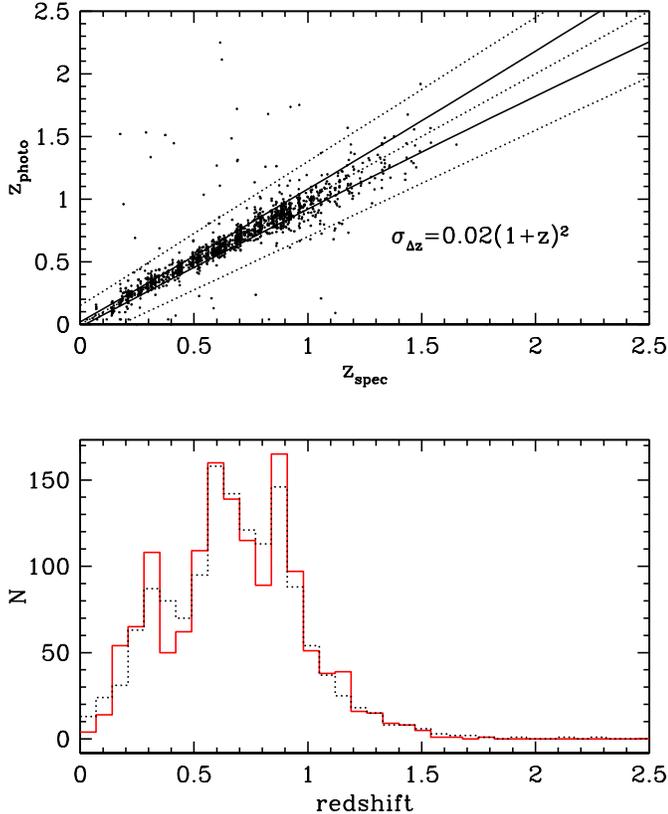}
\caption{ {\it Upper panel:} Comparison of photometric and spectroscopic redshifts in the 
$K$-selected sample for objects
with highly reliable (confidence level $>97\%$ , i.e. about 1400 galaxies with
flag=3, 4) spectroscopic redshifts. The accuracy obtained is 
$\sigma_{\Delta z}=0.02(1+z)^2$ (shown as solid lines) with only 3.7\% of outliers, defined as the objects
outside the region limited by the 2 dotted lines ($z_{photo}=z_{spec}\pm 0.15(1+z_{spec})$)
in the figure. {\it Lower panel:} Spectroscopic (solid line) and photometric (dotted line) redshift 
distribution for the same comparison sample. 
}
\label{fig:zszp}
\end{figure}
%-------------------------------------------------------------------------------

\subsection {The $I$-selected Spectroscopic Sample}\label{sec:Isample}

In this study we use the F02-VVDS deep spectroscopic sample, purely magnitude limited
($17.5\leq I\leq24$), in combination with the multi-wavelength optical/near-IR dataset.
From the total sample of $8281$ objects with measured redshift,
we removed the spectroscopically confirmed stars and broad line AGN, as well
the galaxies with low quality redshift flag (i.e. flag 1),
remaining with 6419 galaxy spectra with secure spectroscopic
measurement (flags 2, 3, 4, 9), corresponding to a confidence level
higher than 80\%. Galaxies with redshift flags 0 and 1 are taken into account
statistically (see Section \ref{sec:GSMF} and Ilbert et al. 2005 and Zucca et
al. 2006 for details).  This spectroscopic sample
has a median redshift of $\sim 0.76$.
Compared to previous optically selected samples, the VVDS has not only
the advantage of having an unprecedented high fraction of spectroscopic redshifts 
(compared, for example, to the purely photometric redshifts as in COMBO17, Wolf et al. 2003
and Borch et al. 2006 for the MF), 
but also of being purely magnitude selected ($17.5<I<24$), differently, for example,
from the DEEP2 (Bundy et al. \cite{bun06} for the MF) survey, which has a colour-colour selection.
Moreover, the VVDS covers an area from 10 to 40 times wider
than the GOODS-MUSIC field (Fontana et al. 2006) and the FORS Deep Field 
(FDF, Drory et al. 2005), respectively.

\subsection {The $K$-selected Photometric Sample}\label{sec:Ksample}

A wide part of the VVDS-02h field (about 623 arcmin$^2$) has been observed also in the near-IR 
(Iovino et al. 2005, Temporin et al. in preparation).
This allows us to build a $K$-selected sample with
a total area of 610 arcmin$^2$ (after excluding low-S/N borders): 442 arcmin$^2$ are 90\% complete
to $K<22.34$, while 168 arcmin$^2$ are 90\% complete to $K<22.84$ (equivalent to
$K_{\rm Vega}=21$).

This sample consists of $11221$ objects, of which $2882$ have VVDS spectroscopy.
In particular, the deep sample ($K<22.84$) consists of $3821$ objects, of which
749 have VVDS spectroscopy, and 596 of them are galaxies with a secure spectroscopic
identification (flags 2, 3, 4, 9).  This latter deep sample is more than one magnitude deeper 
than the samples from the K20 spectroscopic survey (Cimatti et al. 2002) and the
MUNICS survey (Drory et al. 2001). Additionally, the total $K$-selected sample covers an area 
more than 10 times wider than the K20 and the GOODS-CDFS sample used by Drory et al. (2005)
and 4 times wider than the GOODS-MUSIC field (Fontana et al. 2006).

Since the spectroscopic sampling of the $K$-selected sample is less than satisfactory, 
we take advantage of the high quality photometric redshifts ($z_{photo}$).
The method and the calibration are presented and discussed in Ilbert et al. 
(2006). The comparison sample contains 3241 accurate spectroscopic redshifts
(confidence level $>97\%$ , i.e. flag=3, 4) up to $I=24$ 
obtaining 
a global accuracy of $\sigma_{\Delta z / (1+z)}=0.037$ with only 3.7\% of outliers.
Also in the $K$-selected photometric sample the agreement between photometric
and highly reliable spectroscopic redshifts (about 1400) is excellent (Figure \ref{fig:zszp}). 
We note, however, a non-negligible number of catastrophic solutions with $z_{photo}\simgt 1.2$
and $z_{spec}\simlt1$
which could introduce a bias at high-redshift (see also discussion in Section \ref{sec:2samples}). 
Even if we cannot rely on a wide spectroscopic comparison sample at high-$z$,
the number of galaxies with $z_{photo}>1.2$ is 
similar or only slightly higher (about 20\%) than the number of galaxies with $z_{spec}>1.2$ 
(63 vs. 51, see Figure \ref{fig:zszp})
and have very similar fluxes and colors. 
For this reason we do not expect that our results on the mass function and the mass density will be strongly biased by the effect 
of catastrophic redshifts (see Section \ref{sec:GSMF}).
Furthermore, at high-$z$ the dispersion between photometric and spectroscopic redshifts increases,
but not drammatically, to 
 $\sigma_{\Delta z / (1+z)}\simeq0.05, 0.06$ at $z>1.2, 1.4$. Over the whole redshift range it can be represented with
 $\sigma_{\Delta z }\simeq0.02(1+z)^2$ (shown in Figure \ref{fig:zszp}).

For the whole $K$-selected sample, the median errors on photometric redshifts, based on $\chi^2$ statistics, 
are $\sigma_{zphoto}=0.06$ (0.04 at $z<1$, increasing to 0.14 at $z>1.5$).
As expected, there is also an increase of $\sigma_{zphoto}$ for the faintest objects, but this increase is
only about 0.02 in the faintest magnitude bin. As previously noted by Ilbert et al. (2006) the statistical errors 
are consistent with $\sigma_{\Delta z}$ and could be used as an indication of their accuracy.
We will discuss in the  following sections the effects of these uncertainties and conclude that they do not 
affect significantly our conclusions. 

We note, moreover, that the $K$-selected sample selects a different population, 
in particular of Extremely Red Objects (EROs) at $z_{photo}>1$ (see Section \ref{sec:2samples} and Fig. 3), 
compared to
the $I$-selected sample used to calibrate the derived photometric redshift. Actually,
photometric redshifts greater than 0.8-1.0 for the EROs population have been confirmed spectroscopically 
with very low contamination of low-$z$ objects (Cimatti et al. 2002). Moreover, the
near-IR bands are crucial to constrain photometric redshifts in the redshift desert since the $J$-band is 
sensitive to the Balmer break up to $z=2.5$. Indeed Ilbert et al. (2006) obtain for the deep sample at
$K<23$ the most reliable photometric redshifts on this sub-sample with only
2.1\%  of outliers and $\sigma_{\Delta z/(1+z)}=0.035$ (see their figure 13).

In this paper we therefore use 
photometric redshifts for the whole $K$-selected photometric sample
and the highly reliable spectroscopic redshifts when available.

In order to select galaxies from the total $K$-selected photometric sample, we
have used a number of photometric methods to remove candidate stars, as
described below.  Some of the possible criteria to select stars are: 
(i) the CLASS\_STAR parameter given by SExtractor (Bertin \& Arnouts \cite{ber96}), providing the
``stellarity-index'' for each object, reliable up to $I\simeq 21$; 
(ii) the FLUX\_RADIUS\_K parameter, computed by SExtractor from the $K$
band images, which gives an estimate of the radius containing half of
the flux for each object; this can be considered a good criterion to
isolate point-like sources up to $K\simeq 19$ (see Iovino et
al. 2005);
(iii) the $BzK$ criterion, proposed by Daddi et al. (2004a), with stars
characterized by colours $z-K < 0.3(B-z)-0.5$; 
(iv) the $\chi^2$ of the SED fitting carried out during the photometric
redshift estimate (Ilbert et al. 2006), with template SEDs of both
stars and galaxies.  

To efficiently remove stars in the whole magnitude range of our sample, 
avoiding as much as possible to lose galaxies, we decided to use the
intersection of the first three criteria. 
We therefore selected as stars the objects 
fulfilling all the
constraints (i) CLASS\_STAR $\ge 0.95$ for $I<22.5$ or
CLASS\_STAR $\ge 0.90$ if $I>22.5$, (ii) FLUX\_RADIUS\_K $< 3.4$ and 
(iii) $z-K < 0.3(B-z)-0.5$.  
When it was not possible to apply criterion (iii), because of non
detection either in the $B$ or $z$ filters, we used criterion (iv) in its place.
Furthermore, we added to the sample of candidate stars also the
objects with $K<16$ and FLUX\_RADIUS\_K $<4$, to be sure to exclude
from the galaxy sample these saturated point-like objects.
The final sample consists of 653 candidate stars, which we
have removed from the sample in the following analysis. Comparing to the spectroscopic subsample
(we remind that stars were not excluded from the spectroscopic targets of VVDS),
we found about 87\% of efficiency to photometrically select stars, 
i.e. only 28 out of the 214 spectroscopic stars
have not been selected in this way, and only 3 (1.4\%) with
highly reliable spectroscopic flag (3, 4),
whereas 21 spectroscopic extragalactic objects (less than 1\%) 
fall inside the candidate star sample. Three of them are broad line AGN and 
the others are all compact objects, most of them with redshift flags 1 or 2
and only one with flag 3. This latter object has not been eliminated from the
galaxy sample.  We have furthermore removed from the galaxy sample 
the spectroscopically confirmed AGNs and the three secure spectroscopic stars which
were not removed with the photometric criteria. 

The final $K$-selected sample consists of $10160$ galaxies with
either photometric redshifts or highly reliable spectroscopic
redshifts, when available, in the range between $0$ and $2.5$: $6720$
galaxies in the shallow $K<22.34$ area of $442$ arcmin$^2$ and $3440$
galaxies in the deeper area ($K<22.84$) of $168$ arcmin$^2$.

%-------------------------------------------------------------------------------
\begin{figure}
\centering
\includegraphics[width=\hsize]{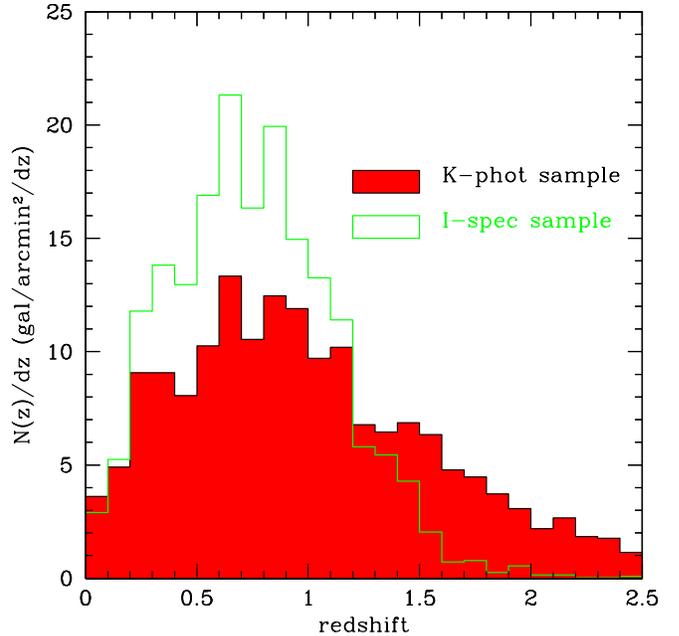}
\caption{Redshift distributions for the $K$-selected photometric 
sample (filled histogram) and for the $I$-selected spectroscopic sample 
(empty histogram).
}
\label{fig:nz}
\end{figure}
%-------------------------------------------------------------------------------

%-------------------------------------------------------------------------------
\begin{figure}
\centering
\includegraphics[width=\hsize]{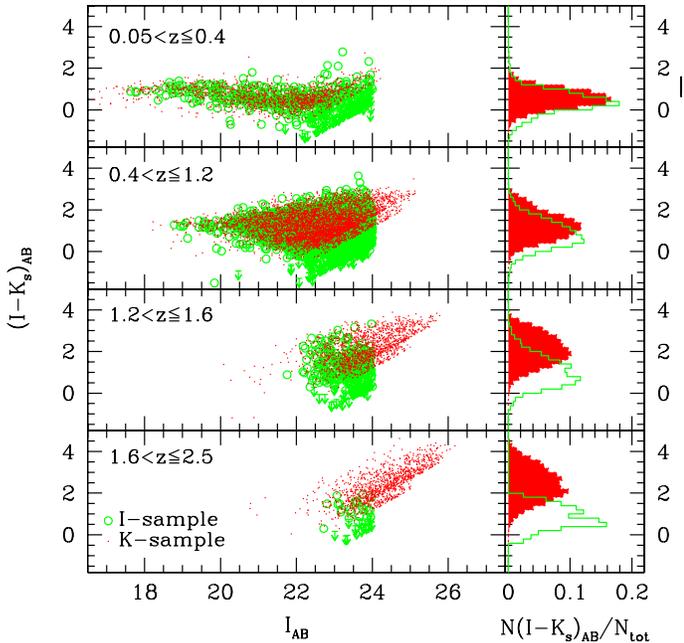}
\caption{Colour-magnitude diagram and colour distribution at different redshifts
for the $K$-selected photometric sample (filled squares and histogram) 
and for the $I$-selected spectroscopic sample (open circles and empty histogram)
}
\label{fig:iik}
\end{figure}
%-------------------------------------------------------------------------------

\subsection{Comparison of the Two Samples}\label{sec:2samples}

As shown in Fig. \ref{fig:nz}, the redshift distribution  in the $K$-selected sample
peaks at higher redshift than in the $I$-selected spectroscopic sample, with the two median
redshifts being 0.91 and 0.76, respectively. 
Even if we cannot rely on a wide spectroscopic comparison sample at high-$z$,
we have better investigated the reliability of the high-$z$ tail in the $K$-selected sample in term of fraction 
and colors. 
We found indeed that at $K<22$ the fraction of objects with $z>1, 1.5$ ($35, 13$\% respectively) is similar
to previous spectroscopic (K20 survey, see Cimatti et al. 2002) or photometric studies (Somerville et
al. 2004).
Moreover, we have used  the $BzK$ color-color diagnostic proposed and calibrated on a spectroscopic sample 
to cull galaxies at $1.4<z<2.5$ (Daddi et al. 2004). 
We found that most (92\%) of the galaxies at $1.5<z_{photo}<2.5$ lie in the high-$z$ region of the $BzK$ diagram.

We conclude that our $K$-selected sample shows no indication of significant bias in its high-redshift tail.
The global $(I-K)$ colour distributions of
the two samples are similar to each other up to $z \sim 1.2$ (see right panels in Fig. \ref{fig:iik}),
but they are significantly different at higher redshift. At $z>1.2$ the $I$-selected sample 
misses many red galaxies fainter than the $I$ limit, most of them being 
Extremely Red Objects (EROs: defined as objects with colours $I-K>2.6$), which are instead included 
in the $K$ sample ($\sim 81$\% of EROs in the deep $K<22.84$ sample have $I>24$). For this reason the 
$K$-selected sample is more adequate to study the massive tail of the GSMF at high-$z$. Vice versa,
the $K$-selected sample misses at all redshifts a number of faint blue galaxies, which are included in the
$I$-selected spectroscopic sample (see left panels in Fig. \ref{fig:iik}). These faint blue
galaxies are important in the estimate of the low-mass tail of the GSMF.

\section{Estimate of the Stellar Masses}\label{sec:mass}

The rest-frame near-IR light has been widely used as a
tracer of the galaxy stellar mass, in particular for local galaxies
(e.g. Gavazzi et al. 1996; Madau, Pozzetti \& Dickinson
1998, Bell \& de Jong 2001). However, an accurate estimate of the galaxy stellar mass at
high $z$, where galaxies are observed at widely different evolutionary stages,
is more uncertain because of the variation of the ${\cal M}_{\rm stars}/L_K$ ratio as a
function of age and other parameters of the stellar population,
such as the star formation history and the metallicity.
The use of multiband imaging from UV to near-IR bands is a way to take
into account the contribution to the observed light of both the old and the young
stellar populations in order to obtain a more reliable estimate of the stellar mass.

However, even stellar masses estimated using the fit to the multicolour spectral
energy distribution (SED) are model dependent (e.g. changing with different assumptions on the
initial mass function, IMF) and subject to various
degeneracies (age -- metallicity -- extinction). In order to reduce such degeneracies
we have used a large grid of stellar population synthesis models, covering a wide
range of parameters, in particular in star formation histories (SFH).  Indeed, in the
case of real galaxies the possibly complex star-formation histories and the presence of
major and/or minor bursts of star formation can affect the derived mass estimate
(see Fontana et al. 2004).

We have applied and compared two different methods to estimate the stellar masses from
the observed magnitudes (using 12 photometric bands from $u^*$ to $K$), 
that are based on different assumptions on the star-formation
history. For both of them we have adopted the Bruzual \& Charlot (2003; BC03 hereafter) code for
spectral synthesis models, in its more recent rendition, using its low
resolution version with the ``Padova 1994'' tracks.
Different models have also been considered, e.g. Maraston 2005,
and P\'egase models (Fioc and Rocca-Volmerange 1997). The results obtained with these
models are compared with those obtained with the BC03 models at the end of Section 
\ref{sec:smoothSFH}.

Since most of previous studies at high-$z$ assumed models with
exponentially decreasing SFHs, we have used the same simple
smooth SFHs (see Section \ref{sec:smoothSFH}) in order to compare our
results with those of previous surveys.  In addition, to further test
the uncertainties in mass determination, we have used models
with complex SFHs (see Section \ref{complexSFH}), in which secondary
bursts have been added to exponentially decreasing SFHs. These
models have been widely used in studies of
SDSS galaxies (see Kauffmann et al. 2003 and Salim et al. 2005 for
further details).  Table \ref{tab:param} summarizes the model parameters
used in the 2 methods described in the following sections.

In our analysis we have adopted the Chabrier IMF (Chabrier et al. 2003),
with lower and upper cutoffs of 0.1 and 100 $M_{\odot}$. 
Indeed, all empirical determinations of the IMF indicate that
its slope flattens below $\sim 0.5\; M_\odot$
(Kroupa \cite{kro01}, Gould et al. \cite{gou96}, Zoccali et al.
\cite{zoc00}) and a similar flattening is required to reproduce the observed
${\cal M}_{\rm stars}/L_B$ ratio in local elliptical galaxies (see e.g. Renzini \cite{ren05}).
As discussed extensively by Bell et al. (2003), the Salpeter 
IMF (Salpeter 1955) is too rich in low mass stars to satisfy dynamical constraints 
(Kauffmann et al. 2003, Kranz et al. 2003). Moreover, di Serego Alighieri et al.
(2005) show a rather good agreement between dynamical masses and stellar masses
estimated with the Chabrier IMF at $z\sim 1$. Specifically, this is true 
at least for high-mass elliptical
galaxies, less affected than lower-mass galaxies by uncertainties in the 
estimate of their dynamical mass due to possibly substantial rotational contribution
to the observed velocity dispersion.

At fixed age the masses obtained with the Chabrier IMF are
smaller by a factor $\sim 1.7$, roughly independent of the age of the
population, than those derived with the classical Salpeter IMF, used in
several previous works that we shall compare with (e.g. Brinchmann \&
Ellis 2000; Cole et al. 2001; Dickinson et al. 2003; Fontana et al. 2004). We have checked this
statement in our sample, finding a systematic median offset of a factor
$1.7$ and a very small dispersion ($\sigma=0.082$ dex) in the masses derived with
the two different IMFs. Since this ratio is approximately constant for a wide range of
star formation histories (SFH), the uncertainty in the IMF does not introduce a
fundamental limitation with respect to the results we will discuss in the following Sections.
Even if the absolute value of the mass estimate is uncertain, the use of Salpeter or Chabrier IMFs
does not introduce any significant difference in the relative evolution with redshift of the mass function
and mass density.

One possible limitation of our approach to derive stellar masses in our sample
is the contamination by narrow-line AGN (broad line AGN have been already excluded,
see Section \ref{sec:Isample}). From the available spectroscopic diagnostic,
in the $I$-selected spectroscopic sample at the mean redshift $z\simeq 0.7$, 
we found that the contamination due to type II AGN is less than 10\%.
Recently, several studies (Papovich et al. 2006, Kriek et al. 2007, Daddi et al. 2007) suggest that
the fraction of type II AGN increases with redshift and stellar mass.
According to Kriek et al. (2007)
at $2<z<2.7$ and $K<21.5$ the fraction is about $20\%$ for massive ($2.9 \times 10^{11} M_\odot$ for
a Salpeter IMF) galaxies. 
To derive the contribution of type II AGN to
the massive tail of the MF is beyond the scope of this paper. However we note, as shown also by the above studies,
that for most of these objects the optical light is dominated by the integrated stellar emission. 
Therefore, both our photometric redshift and mass estimates are likely to be 
approximately correct also for them.

\subsection{Smooth SFHs}\label{sec:smoothSFH}

Consistently with previous studies, we have used synthetic models with
smooth SFH models (exponentially decreasing SFH with time scale $\tau$: 
$SFR (t) \propto \exp(-t/\tau)$) and a best-fit technique to derive stellar masses
from multicolour photometry.

To this purpose we have developed the code {\it HyperZmass}, a modified
version of the public photometric redshift code HyperZ (Bolzonella et
al. 2000): like the public version, {\it HyperZmass} uses the SED fitting technique,
computing the best fit SED by minimizing the $\chi^2$ between observed and
model fluxes.  We used models built with the Bruzual \& Charlot (2003)
synthetic library. 
When the redshift is known, either
spectroscopic or photometric, the best fit SED and its normalization
provide an estimate of the stellar mass contained in the observed
galaxy. In particular, we estimate the stellar mass content of the
galaxies, derived by BC03 code, by integrating the star formation history over the galaxy age
and subtracting from it the ``Return fraction" (R) due to mass loss during
the stellar evolution. 
For a Chabrier IMF, this fraction is already as high
as $\sim 40$\% at an age of the order of 1 Gyr and approaches 
asymptotically about 50\% at older ages.

The parameters used to define the library of synthetic models are
listed in Table \ref{tab:param}. Similar parameters have been used in
Fontana et al. (2004). The Calzetti (2000) extinction law has been used. 
Following that paper, we have excluded 
from the grid some models which may be not physical (e.g. those implying
large dust extinctions, $A_V>0.6$, in absence of a significant star-formation rate, $Age/\tau>4$, see Table \ref{tab:param}).
To better match the ages of early-type galaxies in the local universe and
following SDSS studies, we also removed models with $\tau<1$ Gyr and with
star formation starting at $z<1$.
								  
We find that the ``formal'' typical $1\sigma$ statistical errors 
(defined as the 68\% range as derived from the $\chi^2$ statistics) on
the estimated masses, not taking into account the error on the estimate of the photometric 
redshift for the $K$-selected sample, are of the order of $0.04$ dex for the $K$-selected sample
and 0.05 for the $I$-selected sample. 
A more reliable estimate of the errors
has been obtained using HyperZ to simulate catalogs to the same depth of our
sample (see Bolzonella et al. 2000). Using all 12 photometric bands (from $u^*$ to $K$),
available for a subset of our data, and realistic photometric errors,
the recovered stellar masses reproduce the input masses with no significant offset and 
a dispersion of 0.12 dex up to $z\sim3$. For comparison, using only the
optical bands (from $u^*$ to $z'$) the dispersion increases to $\sim 0.49$ dex at $z>1$.
The best fit masses obtained from input simulations built using randomly all
available metallicities and analyzed only with solar metallicity models are not
significantly shifted from the input masses, but the dispersion increases
from 0.12 dex to 0.21 dex. These dispersions, computed using a $4\sigma$ 
clipping, provide an estimate of the minimum, intrinsic uncertainties
of this method at our depth. 
For the $K$-selected photometric sample further uncertainties in the fitting technique
are due to the photometric redshift accuracy 
($\sigma_{\Delta z}\simeq0.02(1+z)^2$ up to $z=2.5$) 
which corresponds on average to about 0.12 dex of uncertainty in mass, being larger at low 
redshift ($\sim$ 0.2 dex at $z<0.4$) than at high-$z$ ($\sim 0.10$ dex at $z=2$).

Although in principle the best-fitting technique provides estimates also for age, 
metallicity, dust content and SFH timescale, our simulations show that on average
all these quantities are much more affected by degeneracies and therefore less 
constrained than the stellar mass. 

%%%%%%%%%%%%%%%%%%%%%%%%%%%%%%%%%%%%%%%%%%%%%%%%%%%%%%%%%%%%%%%%%%%%%%%%%%%%%%
\begin{table}
\caption[]{Parameters Used for the Library of Template SEDs}
\label{tab:param}
$$ 
\begin{array}{p{0.3\linewidth}ll}
\hline
\hline
\noalign{\smallskip}
Method & Smooth ~SFHs & Complex ~SFHs \\
\noalign{\smallskip}
\hline
\hline
\noalign{\smallskip}
IMF & Chabrier & Chabrier\\
\noalign{\smallskip}
\hline
\noalign{\smallskip}
SFR $\tau$ (Gyr) & [0.1, \infty ]^{\mathrm{a}} & [1, \infty] \\
\noalign{\smallskip}
\hline
\noalign{\smallskip}
log(Age)$^{\mathrm{b}}$ (yr) & [8,10.2]& [8,10] \\
\noalign{\smallskip}
\hline
\noalign{\smallskip}
burst age (yr) &  - & [0,10^{10}] \\
\noalign{\smallskip}
\hline
\noalign{\smallskip}
burst fraction &  - & [0,0.9] \\
\noalign{\smallskip}
\hline
\noalign{\smallskip}
Metallicities & Z_{\odot} & [0.1Z_{\odot},2Z_{\odot}] \\
\noalign{\smallskip}
\hline
\noalign{\smallskip}
Extinction & Calzetti ~law  & Charlot \& Fall ~model \\
& & (n=0.7, \mu \in [0.1,1])  \\
\noalign{\smallskip}
\hline
\noalign{\smallskip}
Dust content& A_V^{\mathrm{c}} \in [0, 2.4]  & \tau_V\in[0,6] \\
\noalign{\smallskip}
\hline
\end{array}
$$ 
\begin{list}{}{}

\item[$^{\mathrm{a}}$] $\tau<1$ if star formation starts at $z<1$.
\item[$^{\mathrm{b}}$] At each redshift, galaxies are forced to have ages smaller than 
the Hubble time at that redshift.
\item[$^{\mathrm{c}}$] $A_V<0.6$ if $Age/\tau>4$.
										\end{list}
\end{table}
%%%%%%%%%%%%%%%%%%%%%%%%%%%%%%%%%%%%%%%%%%%%%%%%%%%%%%%%%%%%%%%%%%%%%%%%%%%%%%

In addition, we have compared our derived masses with those obtained by
using different population synthesis models, such as P\'egase and Maraston
(2005) models. In particular, Maraston (2005) models include the thermally
pulsing asymptotic giant branch (TP-AGB) phase, calibrated with local stellar
populations. This stellar phase is the dominant source of bolometric and 
near-IR energy for a simple stellar population in the age range 0.2 to 2 Gyr. 
We have tested the differences with BC03 models using the $I$-selected spectroscopic
sample and we found only a small but systematic shift ($\sim -0.14$ dex and a similar 
dispersion) up to $z\sim 1.2$ both with and without the use of near-IR photometry. 
On the contrary, masses derived using P\'egase models and similar SFHs have instead no significant 
offset.

At higher redshifts the differences between our estimated masses and those obtained
with Maraston models in the $K$-selected spectroscopic subsample are slightly smaller
and even smaller in the $K$-selected photometric sample ($\sim -0.11,-0.08$, respectively). 
This differences are smaller than that found by Maraston et al. (2006), $\sim-0.2$,
in their
SED fitting (from $B$ up to Spitzer IRAC and MIPS bands) of a few high redshift passive
galaxies
with typical ages in the range 0.5 -- 2.0 Gyr,
selected in the Hubble Ultra Deep Field (HUDF). 
This difference between our results and those of Maraston et al.
could be due to a combination of effects, such as the absence in our photometric
data of mid-IR Spitzer photometry, which at these redshifts is sampling the rest frame
near-IR part of the SED, mostly influenced by the TP-AGB phase, and also 
to the wide range
of complex stellar populations in our sample, in which the effect of the TP-AGB phase may
be diluted by the SFH.

\subsection{Complex SFHs}\label{complexSFH}

Real galaxies could have undergone a more complex SFH, in particular
with the possible presence of bursts of star formation on the top of
a smooth SFH. Thus, we have computed masses also following a different
approach, which has been intensively used in previous studies of SDSS
galaxies (e.g. Kauffmann et al. 2003, Brinchmann et al. 2004, Salim
et al. 2005, Gallazzi et al. 2005). In this approach we parameterize
each SF history in terms of two components: an underlying continuous
model, with an exponentially declining SF law
($SFR(t)\propto\exp(-t/\tau)$), and random bursts superimposed on
it. We assume that random bursts occur with equal probability at
all times up to galaxy age.  They are parameterized in terms of the
ratio between the mass of stars formed in the burst and the total mass
of stars formed by the continuous model over the age.  This
ratio is taken to be distributed between 0.0 and 0.9.  During a burst,
stars are assumed to form at a constant rate for a time distributed
uniformly in the range 30 -- 300~Myr. The burst probability is set so that 50\% of the
galaxies in the library have experienced a burst in the past 2~Gyr.
Attenuation by dust is described by a two-component model (see Charlot
\& Fall 2000), defined by two parameters: the effective $V$-band
absorption optical depth $\tau_V$ affecting stars younger than 10~Myr
and arising from giant molecular clouds and the diffuse ISM, and the
fraction $\mu$ of it contributed by the diffuse ISM, that also affects
older stars. We take $\tau_V$ to be distributed between 0 and 6 with a
broad peak around 1 and $\mu$ to be distributed between 0.1 and 1 with a
broad peak around 0.3. Finally, our model galaxies have metallicities uniformly
distributed between 0.1 and 2 $Z_{\odot}$.

%-------------------------------------------------------------------------------
\begin{figure}
\centering
\includegraphics[width=\hsize]{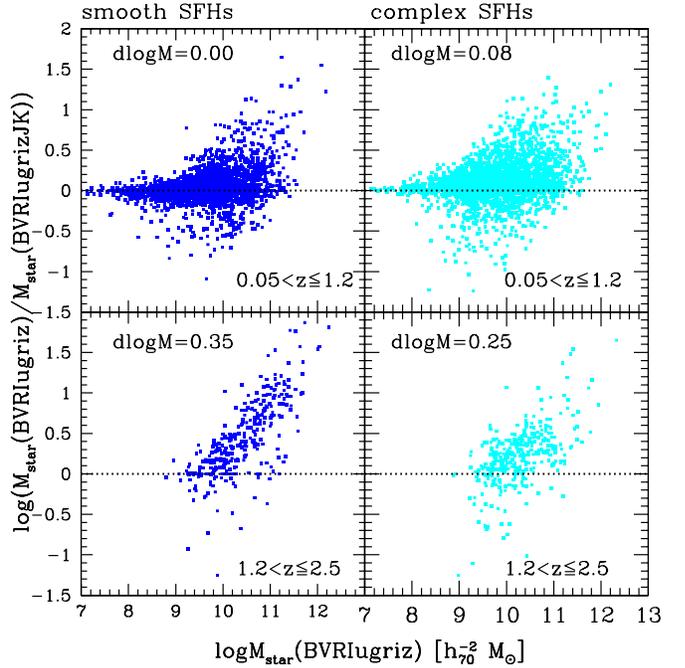}
\caption{Effect of NIR photometry in the mass determination:
ratio between masses estimated without and with NIR photometry vs. 
mass determined without NIR photometry.
The data have been splitted into different redshift ranges.
{\it Left:} masses determined using smooth SFHs.
{\it Right:} The same, but using complex SFHs} 
\label{fig:mmJK}
\end{figure}
%-------------------------------------------------------------------------------

The model spectra are computed at the galaxy redshift and 
in each of them we measure the $k$-shifted model magnitudes for each VVDS
photometric band.  We also force the age of all models in a specific
redshift range to be smaller than the Hubble time at that
redshift. 
The model SEDs are then scaled to each observed SED with a least squares
method and the same scaling factor is applied to the model stellar mass. 
We compare the observed to the model fluxes in each photometric band and
the $\chi^2$ goodness of fit of each model determines the {\it weight} ($\propto
\exp[-\chi^2/2]$) to be assigned to the physical parameters of that
model when building the probability distributions for each parameter
of any given galaxy.  The probability distribution function (PDF) of a
given physical parameter is thus obtained from the distribution of the
weights of {\it all} models in the library at the specified redshift.
We characterize the PDF using its median and the 16 -- 84
percentile range (equivalent to $\pm1\sigma$ range for Gaussian
distributions), and also record the $\chi^2$ of the best-fitting
model. 

Similarly to what has been done for the models with smooth SFH (see Section 
\ref{sec:smoothSFH}), also in this case the stellar mass content of galaxies
is derived by subtracting the return fraction R from the total formed 
stellar mass. We find that the average ``formal'' $1\sigma$ error 
(defined as half of the 16 -- 84 percentile range) on the estimated masses
is of the order of $0.09$ dex for the $K$-selected sample.
The average error increases with redshift from $\sim 0.06$ dex at low
redshift to $\sim 0.11$ dex at $1<z<2$ and decreases with increasing mass 
from $\sim 0.08$ dex for $\log M<10$ to $\sim 0.05$  for $\log M>10$ at $z\simeq0.7$. 
In the $K$-selected sample the photometric redshift accuracy induced a further uncertainty on the mass
of the order of 30\% up to $z>1.5$.
In the $I$-selected sample, where near-IR photometry is not always
available, the typical error on the mass is larger and is of the order of $\sim 0.13$ dex.

\subsection{Effect of NIR Photometry}\label{sec:NIR}

For about half of the sources in the $I$-selected sample only optical
photometry is available. We have therefore used the results obtained for
the $K$-selected spectroscopic subsample to better understand the reliability
of the mass estimates in the whole $I$-selected sample
and quantify potential systematic effects. We found that the 
mass estimates derived using only optical bands are on average in rather
good agreement with those obtained using also NIR bands up to $z\sim1.2$.
In absence of NIR bands the galaxy stellar masses tend to be
only slightly overestimated, with a median shift $<0.1$ dex;
this is due to the fact that already at $z=0.4$, for example, 
the $z'$-band (the reddest band used in the fit in absence of NIR) samples the 
$R$-band rest-frame and therefore the SED fitting is less reliable 
for the estimate of the stellar masses.
There is however a significant fraction of the
galaxies for which the ratio between the two masses is higher than a factor
of three (see upper panels of Fig.\ref{fig:mmJK}). This fraction of galaxies
with significantly discrepant mass estimates is $\sim$ 5\% for the models
with smooth SFH and $\sim 9$ \% for the models with complex SFH.

At higher redshifts, where our reddest optical band, i.e. the $z$-band, is
sampling the rest-frame spectrum bluewards of the 4000 \AA\ break, 
the comparison of the two sets of mass estimates (i.e. with and without
near-IR photometry) is significantly worse. Not only the median shift increases
significantly, but also the ratio of the two sets of masses is significantly
correlated with the mass derived without using NIR photometry (see lower panels of Fig. 
\ref{fig:mmJK}). For this reason, we have decided to use the whole $I$-selected
VVDS spectroscopic sample only up to $z\sim1.2$, whereas at higher redshifts
we use as reference the $K$-selected photometric sample. 

As shown in the upper panels of Fig.\ref{fig:mmJK}, the ratio between the masses
computed without and with NIR photometry has a non-negligible dispersion also
for $z < 1.2$, with the masses computed without NIR photometry being higher on average. 
In order to statistically correct for this effect, we have performed the following
Monte Carlo simulation. For each galaxy without near-IR photometry in the $I$-selected
spectroscopic sample we have applied a correction factor to its estimated mass.
This correction factor has been derived randomly from the observed distribution,
at the mass of each galaxy, of the ratios of the masses with and without NIR photometry.
The effect on the mass function of using these ``statistically corrected'' masses is shown
and discussed in Sect. \ref{sec:GSMF}. 

%-------------------------------------------------------------------------------
\begin{figure}
\centering
\includegraphics[width=\hsize]{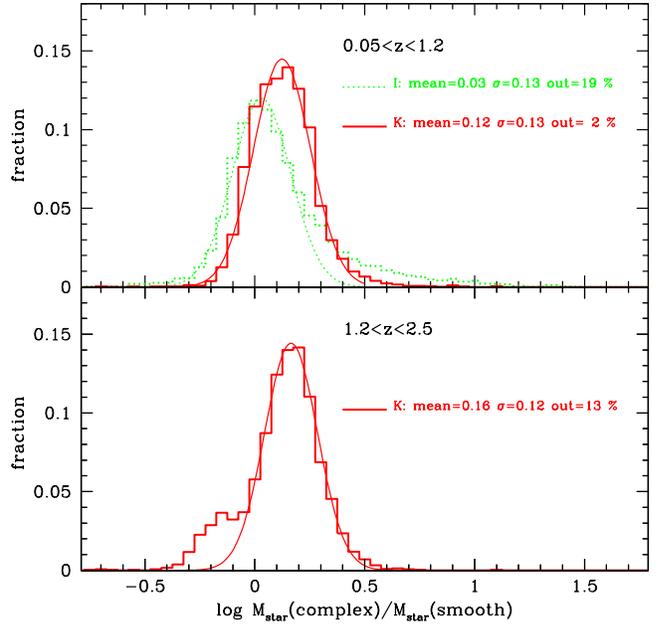}
\caption{Histograms of the ratio of the masses derived with the smooth and
the complex star formation histories. In the upper panel ($z<1.2$) both data
from the $I$-selected and the $K$-selected samples are shown; in the lower panel
($z>1.2$) only data from the $K$-selected sample are shown, since 
the $I$-selected sample is not used to derive the mass function in this redshift range.
}
\label{fig:massmass}
\end{figure}
%-------------------------------------------------------------------------------

\subsection{Comparison of the Masses Obtained with the Two Methods}\label{sec:comp}

In this section we compare the mass estimates we obtained 
using the two different methods described above for the VVDS galaxies.
Since the ratio of the two estimates is almost independent of the
mass, in   Fig. \ref{fig:massmass} we show the histograms of this ratio,
integrated over all masses, for two different redshift bins. In the upper
panel ($z<1.2$), both data from the $I$-selected and the $K$-selected
samples are shown; in the lower panel only data from the $K$-selected
sample are shown, since the $I$-selected sample is not used to derive
the mass function in this redshift range. Gaussian curves, representing
the bulk of the population, are drawn on the top of each histogram.
The parameters of each Gaussian are reported in the figure.

The global comparison of the two sets of masses is rather satisfactory,
even if it shows a systematic shift, 
larger for the $K$-selected sample, between the two sets of masses,
with the `smooth SFH' masses being on average smaller than the 
`complex SFH' masses. The values of $\sigma$ of these Gaussians are similar,
of the order of $0.13$ dex. However, in two of the three cases
(i.e. the $I$-selected sample at low redshift and the $K$-selected sample
at high redshift) the distributions of the ratio of the masses appear
to be asymmetric, with tails which are not well represented by a
Gaussian distribution. The fractions of these ``outliers'' are given
in Fig. \ref{fig:massmass}. This tail is particularly significant for
the $I$-selected sample, for which there are galaxies with the ratio
between the two masses higher than 3 and in a few cases
reaching a value of 10.

We have analyzed the effect of the different parameters used in the two
methods. 
The impact of different extinction curves on the mass
estimates has already been investigated by Papovich et al. (2001),
Dickinson et al. (2003), Fontana et al. (2004), and found to be small.  
We have repeated the same
exercise for the two different dust attenuation models adopted, finding that 
for a given SFH the mass estimated with different extinction laws are similar, with
an average shift of 0.02 dex.

Analyzing in some detail the properties of the galaxies which are in the extended tail
of large mass ratios for the $I$-selected sample at low redshift (see the upper panel in Fig. \ref{fig:massmass}), 
we found that,
even if many of them do not have near-IR photometry and therefore their mass is more uncertain
(see previous section), 
on average they are characterized by blue colours in the bluer bands ($B-I$) and red colours in the
redder bands ($I-z$ or $I-K$). 
Because of this, they have been fitted typically with low values of the
population age by HyperZmass with a smooth SFH, while the estimate obtained with
a complex SFH corresponds to an older, and therefore more massive, population plus a
more recent burst with a mass fraction in the burst of the order of ${\it fburst}<0.15$.
Vice versa the low ratio outliers in the $K$-selected sample at high redshift
have been fitted typically with moderately higher dust content by HyperZmass
than with complex SFHs.
Finally, we conclude that the main differences between the two methods to determine the
masses are largely due to different assumed SFHs and, in particular, to the secondary
burst component allowed in the model with complex SFHs.

To summarize, we have explored in detail two different methods 
and a wide parameter space (see Table \ref{tab:param}) to estimate the
stellar mass content in galaxies in order to better understand the 
uncertainties in the photometric stellar mass determinations. 
Indeed, a good estimate of the intrinsic errors may be critical for the GSMF 
measurement and interpretation. We found that, within a given assumption on the
SFH, the accuracy of the photometric stellar mass is overall satisfactory,
with intrinsic uncertainties in the fitting technique of the order of 
$\sim$ 30\%,
in agreement with similar
results in the K20 (Fontana et al. 2004), HDFN (Dickinson et al. 2003) and 
HDFS (Fontana et al. 2003) at the same redshifts. 
These errors are smaller than the estimates at higher redshift 
($z\simeq 3$) (\cite{Papovich2001}, \cite{shapley}), since at our average
redshifts ($z\simeq 0.7-1$) we can rely on a better sampling of the rest-frame near-IR part
of the spectrum, while are similar to the uncertainties estimated at high-$z$ using IRAC data 
by Shapley et al. (2005). 
For the $K$-selected photometric sample the uncertainties in the stellar mass due to 
the photometric redshift errors are on average of the order of 30\% up to $z>1.5$, giving
a total fitting uncertainty up to 45\% at $z>1.5$.
Finally, systematic shifts, mainly due to different assumptions on the SFHs,
can be as large as $\sim$ 40\% 
over the entire redshift range $0.05<z<2.5$ when NIR photometry is available. 
The uncertainty in the derived masses is obviously higher also at low redshift
when NIR photometry is not available and in this case it becomes extremely large at $z>1.2$ (see
Fig.\ref{fig:mmJK}).

For what concerns the absolute value of the mass, its uncertainty is
mainly due to the assumptions on the IMF and it is within a factor of 2 for the
typical IMFs usually adopted in the literature.

In the following sections we discuss in some detail the effects of the two
methods on the derivation of the GSMF.

\subsection{Massive Galaxies at $z>1$}\label{sec:massive}

\begin{figure}
\centering
\includegraphics[width=\hsize]{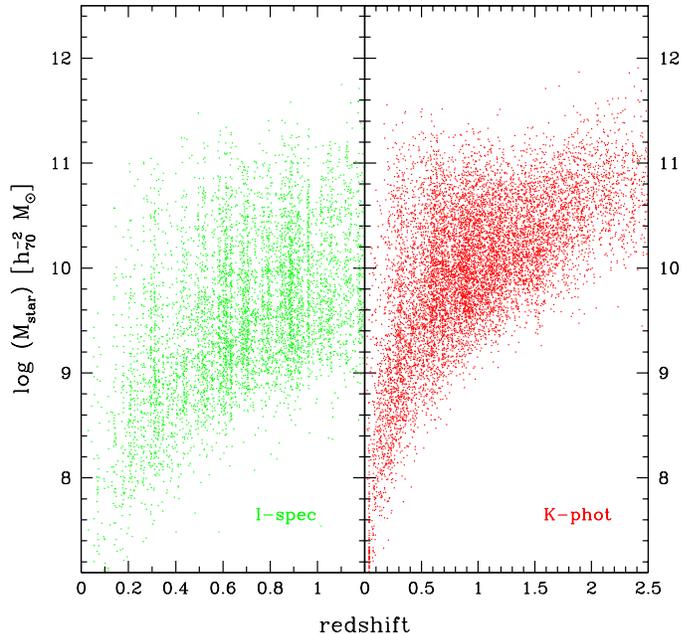}
\caption{Stellar Mass as a function of redshift for the 
$I$-selected spectroscopic (left) and for the $K$-selected photometric (right)
samples for smooth SFHs. 
}
\label{fig:Mz}
\end{figure}

Figure \ref{fig:Mz} shows the stellar masses for the 2 samples ($I$ and 
$K$-selected) derived using the smooth SFHs.
It is interesting to note the presence
of numerous massive objects ($\log M > 11$) at all redshifts and up to $z=2.5$.
High-$z$ massive galaxies have been already observed in previous
surveys (Fontana et al. 2004, Saracco et al. 2005, Cimatti et al. 2004, Glazebrook et al 2004,
Fontana et al. 2006, Trujillo et al. 2006).
Here the relatively wide area  (the $K$-selected sample is more than 10 times wider 
and from 0.5 to 1 magnitude deeper than the K20 survey)
allows to better sample the massive tail of the
population.
We note that massive galaxies have typically redder optical-NIR 
rest-frame colours ($\langle M_I-M_K\rangle\simeq 0.7$) compared to the whole 
population ($\langle M_I-M_K\rangle\simeq0.5$),
consistently with the idea that massive galaxies host the oldest stellar 
population.
Further analysis of the stellar population properties and spectral features 
of massive galaxies, as well as of red objects will be presented in forthcoming 
papers (Lamareille et al. in preparation, Vergani et al. 2007, 
Temporin et al. in preparation). 

%-------------------------------------------------------------------------------
%-------------------------------------------------------------------------------

%-------------------------------------------------------------------------------
\begin{figure*}
\centering
\includegraphics[width=\hsize]{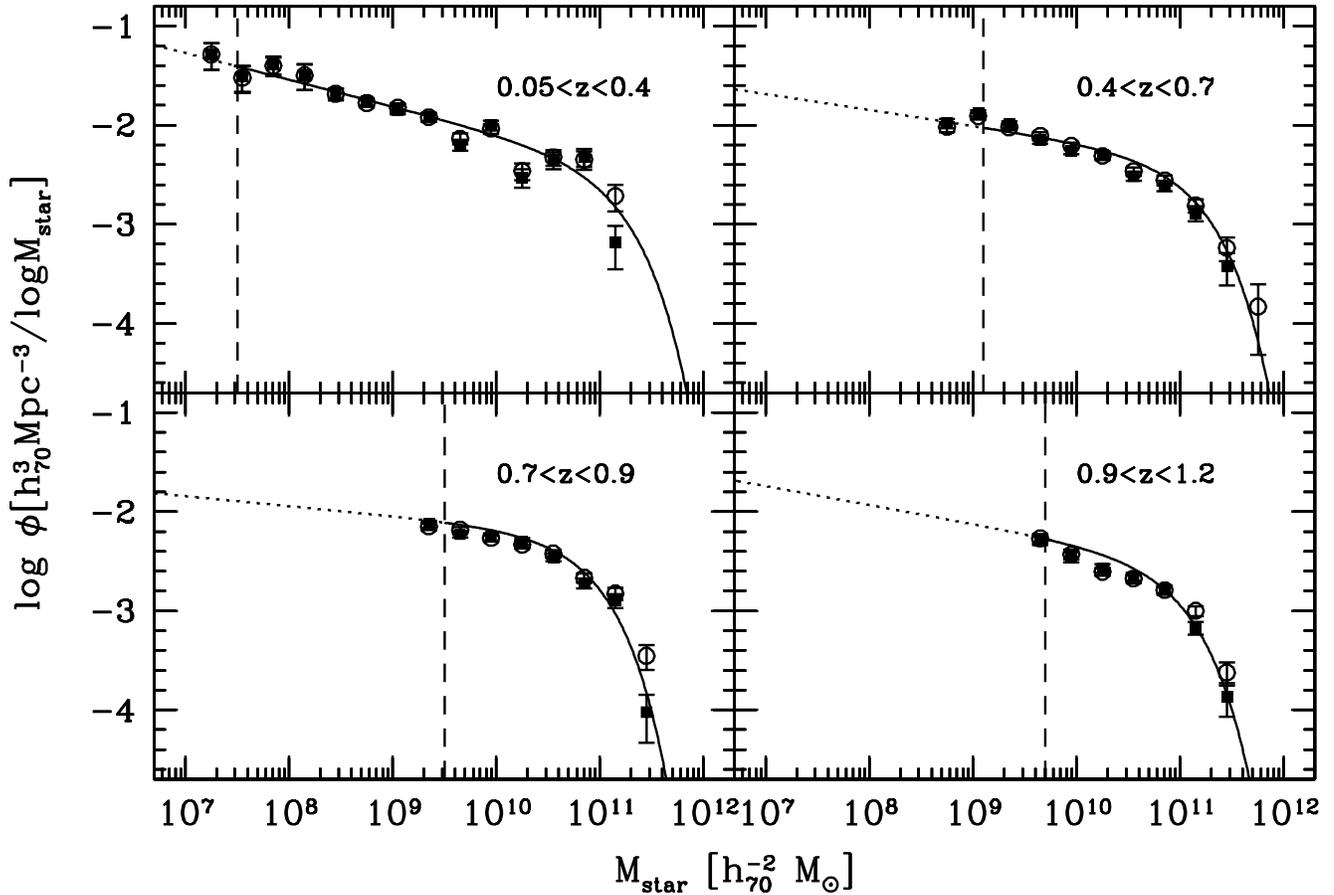}
\caption{
Effect on the $I$-selected MF of the use of the statistically corrected masses,
which take into account the effect of 
near-IR photometry on the mass determination (see text for details).
Complex SFHs have been used to derive masses.
Empty circles represent the MF obtained using the original uncorrected masses, 
filled squares show the MF obtained using the statistically 
corrected masses. 
For comparison the STY Schechter MF is shown for the subsample where 
near-IR photometry is available.
The vertical dashed lines represent the completeness limit of the sample as
defined in the text.
}
\label{fig:MF_simul}
\end{figure*}
%-------------------------------------------------------------------------------
%-------------------------------------------------------------------------------
\begin{figure*}
\centering
\includegraphics[width=\hsize]{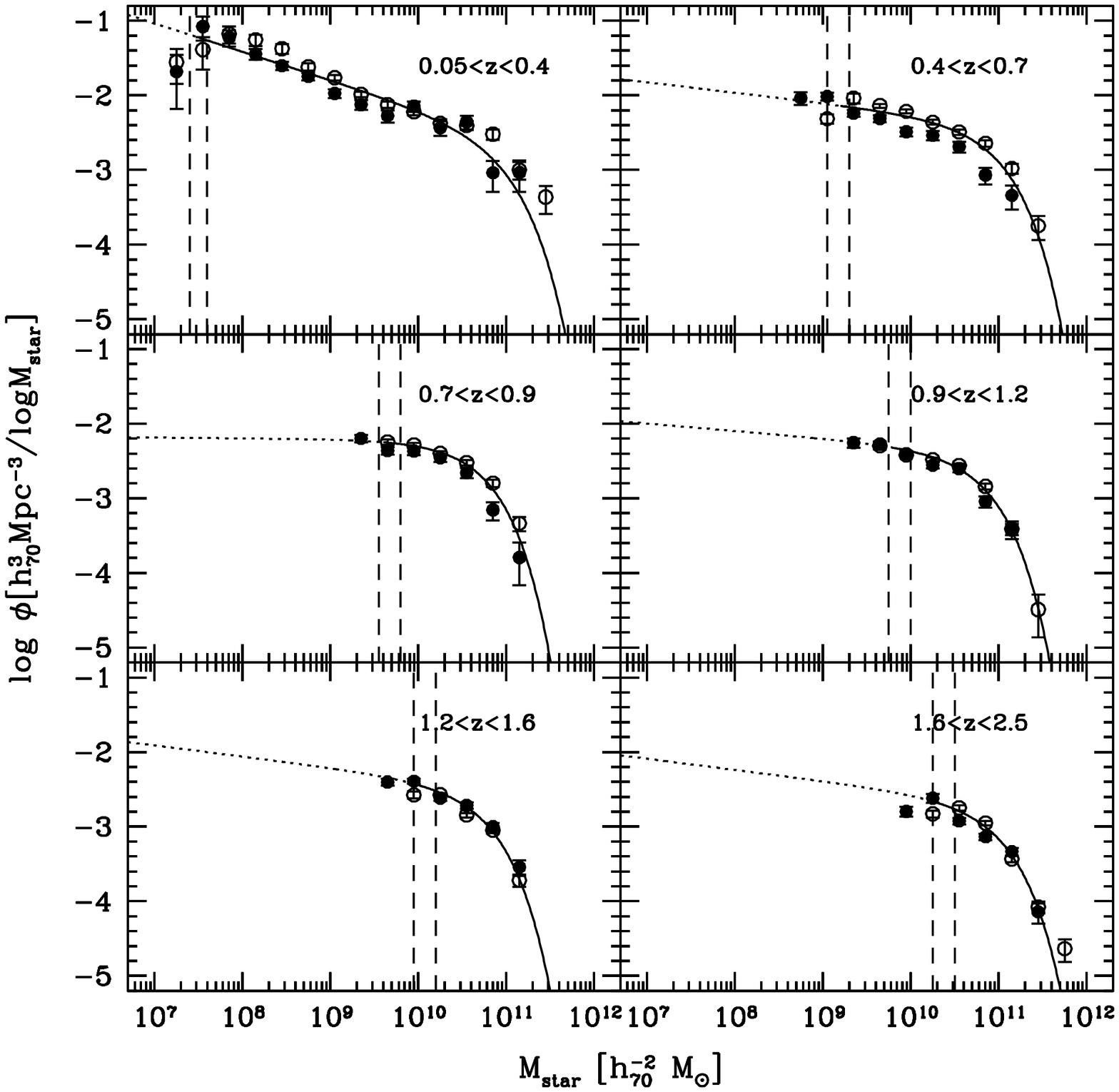}
\caption{$K$-selected MF derived from the 2 subsamples,
deep (filled circles) and shallow (empty circles), separately
(smooth SFHs have been used to derive masses).
For comparison the STY Schechter MF for the global $K$-selected sample is shown.
The vertical dashed lines represent the completeness limit of the 2 $K$-selected subsamples.
}
\label{fig:MF_shallow_deepK}
\end{figure*}
%-------------------------------------------------------------------------------

\section{Mass Function Estimate}\label{sec:GSMF}

Once the stellar mass has been estimated for each galaxy in the
sample, the derivation of the corresponding Galaxy Stellar Mass Function
(MF) follows the traditional techniques used for the computation of the
luminosity function. Here we apply both the classical non-parametric $1/V_{\rm max}$ formalism
(Schmidt 1968, Felten 1976) and the parametric STY (Sandage, Tammann \& Yahil 1979) 
method to estimate best-fit Schechter (1976) parameters ($\alpha, {\cal M}_{\rm stars}^*, \phi^*$). 
In the case of the $K$-selected sample,
in order to take into account the two different magnitude limits,
we perform a ``Coherent Analysis of independent samples" as described by Avni \& Bahcall (1980).

Ilbert et al. (\cite{ilbert04}) have shown that the estimate of the
faint end of the global luminosity function can be biased, because, due
to different k-corrections, different galaxy types have different 
absolute magnitude limits for the same apparent magnitude limit. The
same bias is present also for the low mass end of the mass function.
This is due to the fact that, because of the existing dispersion in the
mass-to-light ratio of different galaxy types, at small masses the objects
with the largest mass-to-light ratio are not included in a magnitude limited
sample
(see Appendix B in Fontana et al. 2004 for an extensive discussion).
For this reason, when computing the global mass functions with the STY method, 
in order to fully avoid this bias, we should use in
each redshift range only galaxies above the stellar mass limit where
all the SEDs are potentially observable. In both the $K$- and $I$-selected samples
these limits derive from the conversion between photometry and 
stellar mass 
for early-type galaxies and are very restrictive. However, recent results
show that the faint-end and the low-mass end of the luminosity and mass function 
is dominated by late-type galaxies 
up to $z\sim 2$ (Fontana et al. 2004, Zucca et al. 2006, Bundy et al. 2006). Therefore for the STY estimate
we will use as lower limit of the mass range the minimum mass above which
late-type SEDs (defined by rest-frame optical/NIR colours $M_I-M_K < 0.4$)
are potentially observable 
(see Fig. \ref{fig:MF}). 

We have estimated the MF for both the deep $I$-selected 
($17.5<I<24$) spectroscopic sample 
and the photometric $K$-selected ($K<22.34$ \& $K<22.84$) sample.
For each sample the MF has been estimated using masses computed with both
methods described in Section \ref{sec:2samples}.
In the case of the spectroscopic sample, 
in order to correct for both the non-targeted sources in spectroscopy and those for which the 
spectroscopic measurement failed, 
we use a statistical weight
$w_i$, associated with each galaxy $i$ with a secure redshift
measurement (see Ilbert et al. \cite{ilb05} for details).  This weight
is the inverse of the product of the {\it Target Sampling Rate}
times the {\it Spectroscopic Success Rate}.  
Accurate weights have been
derived by Ilbert et al. (2005) for all objects with secure spectroscopic
redshifts, taking into account all the parameters involved
(magnitudes, galaxy size and redshift).

For the $K$-selected sample, we have tested the effect of catastrophic photometric redshifts
(see discussion in Sec. \ref{sec:Ksample})
on the evolution of the mass function and mass density.
We have used the $I$-selected spectroscopic sample, replacing spectroscopic redshifts with photometric redshifts.
The two MFs (with either spectroscopic or photometric redshifts) are very similar in the whole mass and redshift range ($0.05<z<2.5$) analyzed and even at $z>1.2$,
where we note a not negligible number of catastrophic photometric redshifts (see discussion in Sec. \ref{sec:Ksample}).
There is no evidence of a strong bias in the normalization and in the shape of the MF; also the massive tails of the MFs are similar, 
within the statistical errors. We conclude that the 
catastrophic solutions at high photometric redshifts (i.e. masses) do not strongly affect our results.

\subsection{The VVDS Galaxy Stellar Mass Function}
%\label{sec:evol}

The resulting stellar mass functions of the VVDS sample are derived 
in the following redshift ranges:
(a) $0.05<z<1.2$ for the $I$-selected sample, because  
at higher redshift the mass estimate becomes very uncertain 
(see figure \ref{fig:mmJK}) and (b) $0.05<z<2.5$ for the 
$K$-selected sample. 
We have furthermore divided the 2 samples into different redshift bins in order 
to sample evolution with similar numbers of sources in each bin. 

%-------------------------------------------------------------------------------
\begin{table*}
\caption[]{STY parameters in the different redshift ranges }
\begin{flushleft}
\begin{tabular}{c c c c c c c} \hline
\hline

%\multicolumn{7} {c}
%{$H_0 = 70$ \hspace{0.3cm} $\Omega_m=0.3$ \hspace{0.3cm}               $\Omega_\Lambda=0.7$} \\
\hline \\
sample & method & $z$ range & mean redshift & $\alpha$ &  $\log {\cal M}_{\rm stars}^* (h_{70}^{-2} M_\odot)$ & $\phi^*$($10^{-3} h_{70}^3 Mpc^{-3}$)   \vspace{0.2cm} \\ \hline
\hline
I & smooth   &  0.05  - 0.4  & 0.27  &  -1.26$^{+0.01}_{-0.02 }$ &  11                     &  1.90$^{+0.08}_{-0.16}$ \\
I & smooth   &  0.4  - 0.7  & 0.58   &  -1.23$^{+0.04}_{-0.04 }$ &  11$^{+0.08}_{-0.08}$   &  1.72$^{+0.33}_{-0.28}$ \\
I & smooth   &  0.7  - 0.9  & 0.8    &  -1.23$^{+0.12}_{-0.11 }$ &  10.88$^{+0.15}_{-0.13}$ &    1.6$^{+0.55}_{-0.45}$ \\
I & smooth   &  0.9  - 1.2  & 1.05   &  -1.09$^{+0.19}_{-0.17 }$ &  10.85$^{+0.14}_{-0.14}$ &    1.3$^{+0.43}_{-0.46}$ \\
\hline
I & complex  &  0.05 - 0.4  & 0.27   &   -1.28$^{+0.02}_{-0.01}$ &  11.15                  &   1.75$^{+0.16}_{-0.08}$ \\
I & complex  &  0.4 - 0.7  &  0.58   &   -1.22$^{+0.04}_{-0.04}$ &  11.15$^{+0.08}_{-0.08}$ &   1.58$^{+0.30}_{-0.26}$ \\
I & complex  &  0.7 - 0.9  &  0.81   &   -1.04$^{+0.08}_{-0.07}$ &  10.83$^{+0.07}_{-0.07}$ &   3.02$^{+0.57}_{-0.51}$ \\
I & complex  &  0.9 - 1.2  &  1.04   &   -1.16$^{+0.1}_{-0.09}$  &  10.89$^{+0.08}_{-0.07}$  &   1.80$^{+0.44}_{-0.39}$ \\
\hline
K & smooth   &  0.05  - 0.4  & 0.26  &  -1.38$^{+0.02}_{-0.01}$  &  10.93                   &  1.29$^{+0.10}_{-0.05}$ \\
K & smooth   &  0.4  - 0.7  &  0.57  &  -1.14$^{+0.04}_{-0.04}$  &  10.93$^{+0.06}_{-0.06}$ &  1.83$^{+0.27}_{-0.24}$ \\
K & smooth   &  0.7  - 0.9  &  0.81  &  -1.01$^{+0.07}_{-0.08}$  &  10.67$^{+0.07}_{-0.05}$ &    2.6$^{+0.38}_{-0.44}$ \\
K & smooth   &  0.9  - 1.2  &  1.05  &   -1.1$^{+0.07}_{-0.08}$  &  10.78$^{+0.06}_{-0.05}$ &   1.83$^{+0.28}_{-0.30}$ \\
K & smooth   &  1.2  - 1.6  &  1.4   &  -1.15$^{+0.12}_{-0.12}$  &  10.72$^{+0.07}_{-0.06}$ &  1.48$^{+0.30}_{-0.30}$ \\
K & smooth   &  1.6  - 2.5  &  1.96  &  -1.15                    &  10.96$^{+0.01}_{-0.02}$ &    0.9$^{+0.30}_{-0.30}$  \\
\hline
K & complex  &  0.05 - 0.4  &  0.26  &   -1.39$^{+0.01}_{-0.02}$ &  11.12                   &   1.17$^{+0.05}_{-0.09}$ \\
K & complex  &  0.4 - 0.7  &   0.57  &   -1.16$^{+0.04}_{-0.04}$ &  11.12$^{+0.06}_{-0.06}$ &   1.58$^{+0.24}_{-0.22}$ \\
K & complex  &  0.7 - 0.9  &   0.81  &   -1.16$^{+0.07}_{-0.07}$ &  10.98$^{+0.07}_{-0.07}$ &   1.74$^{+0.36}_{-0.30}$ \\
K & complex  &  0.9 - 1.2  &   1.05  &    -1.2$^{+0.07}_{-0.06}$ &  11.07$^{+0.06}_{-0.06}$ &   1.34$^{+0.26}_{-0.21}$ \\
K & complex  &  1.2 - 1.6  &   1.4   &   -1.17$^{+0.12}_{-0.12}$ &  10.93$^{+0.07}_{-0.06}$ &   1.39$^{+0.29}_{-0.28}$ \\
K & complex  &  1.6 - 2.5  &   1.96  &   -1.17                   &  10.97$^{+0.01}_{-0.02}$ &   1.25$^{+0.09}_{-0.04}$ \\
\hline
\end{tabular}
\end{flushleft}
\label{tab:MFP}
\end{table*}
%-------------------------------------------------------------------------------

%-------------------------------------------------------------------------------
\begin{figure*}
\centering
\includegraphics[width=0.95\hsize]{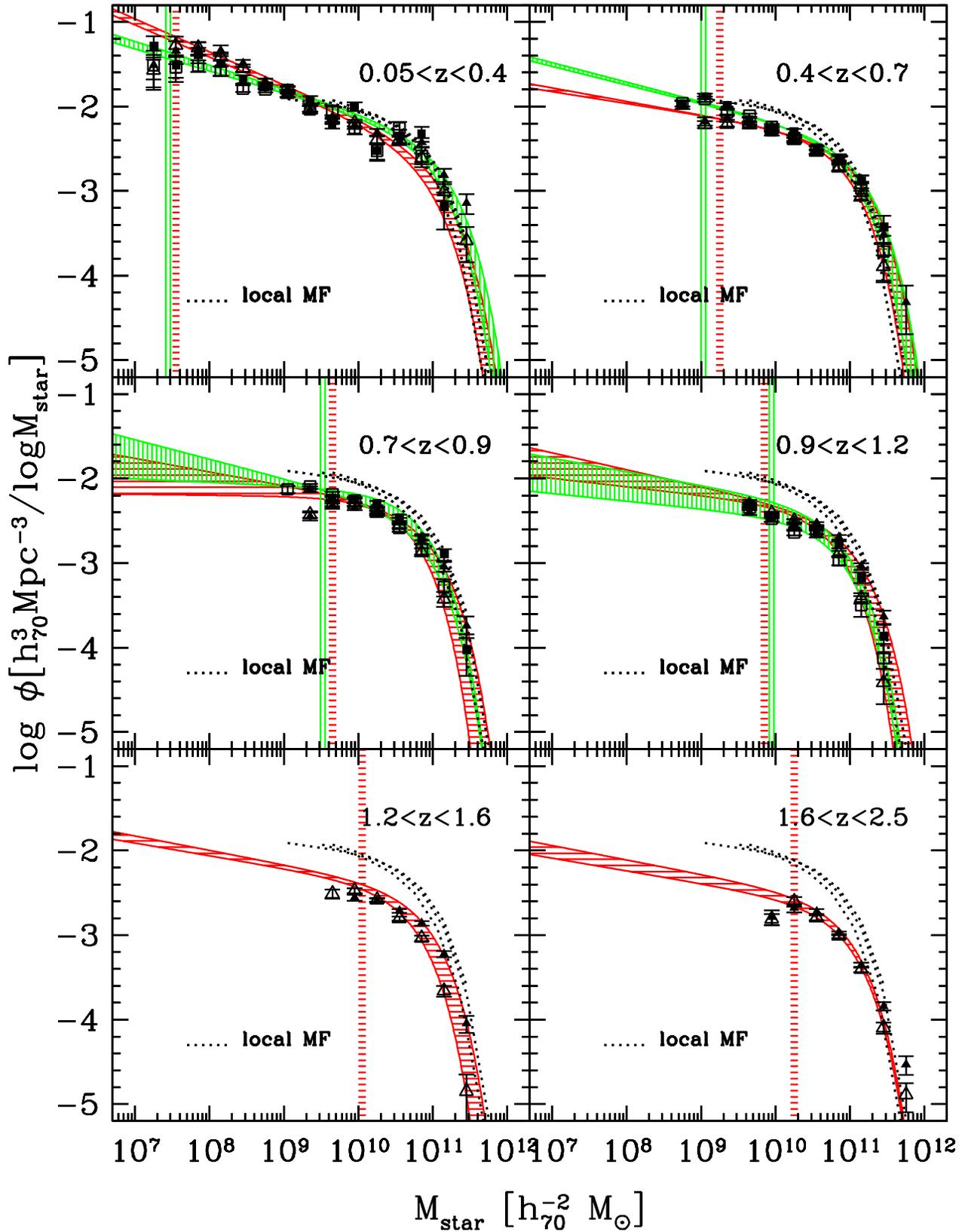}
\caption{Galaxy Stellar Mass Functions in the $I$-selected (squares) 
and $K$-selected (triangles) 
using both methods to estimate the stellar masses
(empty symbols for smooth SFHs and filled for complex SFHs). 
The STY Schechter fits for the 2 methods limit 
the hatched regions (horizontal hatched for the $K$-selected and vertical hatched
for the $I$-selected samples).
Vertical hatched regions represents the completeness limit of the 2 samples.
The local MFs by Cole et al. (2001), both original and ``rescaled" version (Fontana et al. 2004),
and by  Bell et al. (2003) are reported in each panel as dotted lines. 
}
\label{fig:MF}
\end{figure*}
%-------------------------------------------------------------------------------
\begin{figure*}
\centering
\includegraphics[width=\hsize]{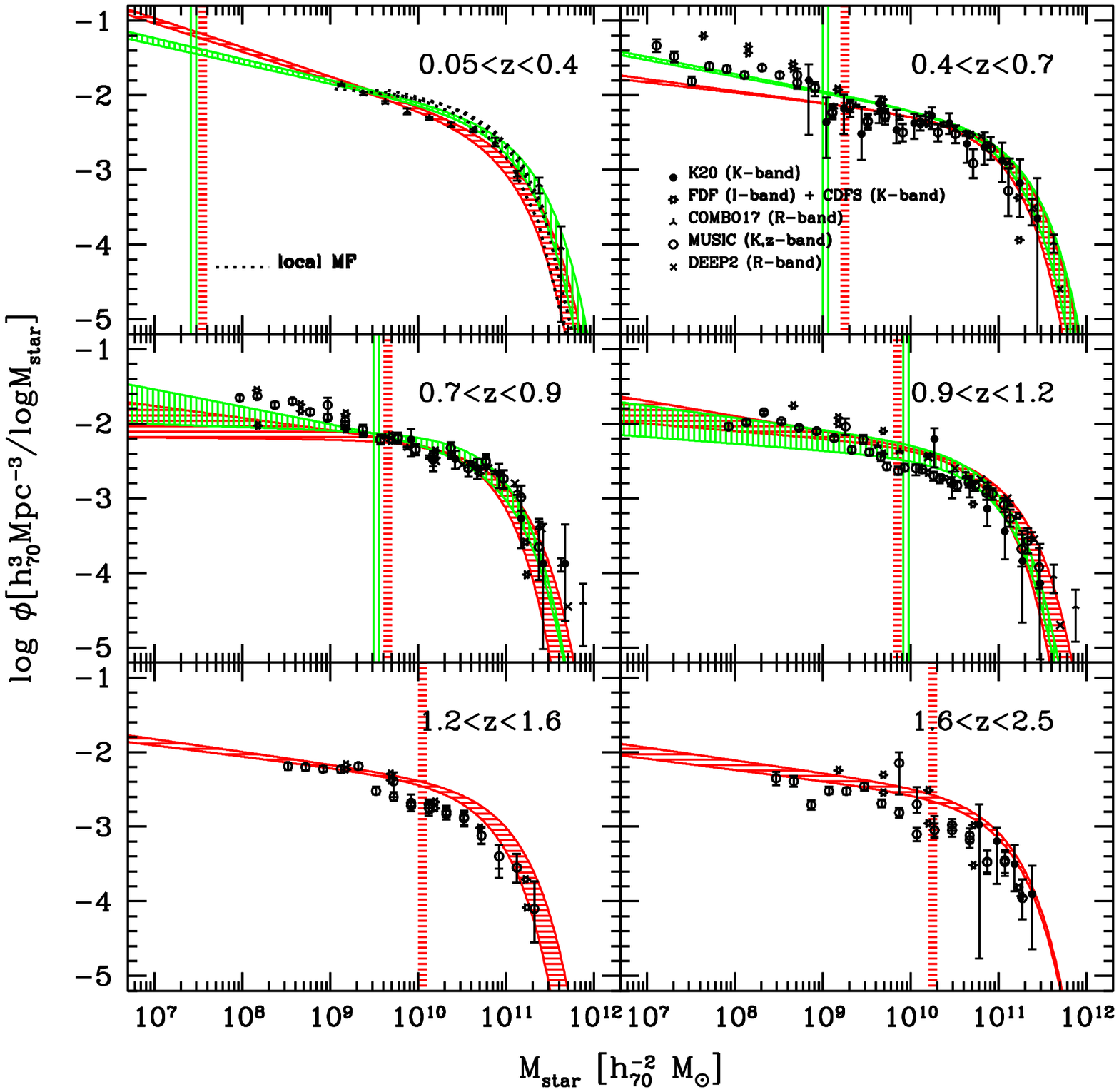}
\caption{Comparison between the $I$-selected and the
$K$-selected MFs in the VVDS (hatched vertical and horizontal STY regions,
respectively, see caption Figure \ref{fig:MF}) and the literature data (K20, COMBO17, MUSIC, DEEP2, FDF+CDFS; 
for each the band of selection is indicated in the parenthesis).
The vertical hatched regions represent the completeness limits of the VVDS
samples.
}
\label{fig:MFlit}
\end{figure*}
%-------------------------------------------------------------------------------

For the galaxies in the $I$-selected sample not covered by near-IR data, 
we have used the statistically corrected masses 
derived through a Monte Carlo simulation,
to take into account the effect of the near-IR photometry in the
mass determination (see Sec. \ref{sec:NIR}).
Figure \ref{fig:MF_simul} shows the effect in the MF for 
complex SFHs. The high mass tail is significantly reduced if we
use statistically corrected masses when near-IR is not available.
Consistent MFs have been  obtained in the sub-area of $I$-selected sample
where near-IR photometry is available (see Fig. \ref{fig:MF_simul}).

For the $K$-selected sample, we have  analyzed the effect of the cosmic variance 
on small areas, 
deriving the MF for the two $K$-selected subsamples, deep and shallow, separately
(see Figure \ref{fig:MF_shallow_deepK}).
We find a significantly lower MF (by a factor 1.8 and 1.6) in the redshift range $0.4<z<0.7$
and $0.7<z<0.9$
in the deep $K$-selected sample ($K<22.84$ over 168 arcmin$^2$) 
compared to the shallow $K$-band sample ($K<22.34$ over 442 arcmin$^2$). 
The significance of such differences in the MF,
is of $\sim 2-3\sigma$ in each mass bin at $0.4<z<0.7$ and $\sim 1-2\sigma$ at $0.7<z<0.9$.
Globally, i.e. for the total number densities over the complete mass range, the differences are
significant at about 5-3 $\sigma$ level in the two redshift ranges, respectively.
This problem leads to a clear warning on the results based on small fields, 
as covered by most of the previous existing surveys.
 
In Figure \ref{fig:MF} we show the MFs derived using the 
$I$-selected spectroscopic sample and the $K$-selected photometric sample
for both methods (smooth and complex SFHs) to derive the masses.
The resulting mass functions are
quite well fitted by Schechter functions.
The best-fit Schechter parameters are summarized in
Table \ref{tab:MFP}, with the uncertainties derived from the projection of
the 68\% confidence ellipse. Since in the lowest and highest redshift 
bins ($z\simeq0.2$ and $2$) the values of ${\cal M}_{\rm stars}^*$ and the low-mass-end slope ($\alpha$), respectively, are
poorly constrained, they have been fixed to the values measured in the following and previous 
redshift bins respectively.

We note, first, that the overall agreement between the MF
derived with the different methods for masses determination is
fairly satisfactory, albeit complex SFHs estimates provide typically
larger masses. 
The systematic shift between the 2 methods 
(Section \ref{sec:comp})
is reflected in most of the redshift bins also in the characteristic mass (${\cal M}_{\rm stars}^*$) of the MF
while the best fit slopes ($\alpha$) and $\phi^*$ Schechter parameters 
agree within the errors between the different methods in most of the redshift bins.

We note, furthermore, an overall agreement between the 2 samples 
($I$- and $K$-selected),  and 
in most of the redshift bins the Schechter parameters agree within the errors, 
even if some differences exist.
More in detail, the $I$-selected sample in the range $0.4<z<0.9$ has a
higher low-mass ($<9.5$ dex) end
and a slightly steeper MF ($\alpha\sim-1.23$) 
than the $K$-selected one ($\alpha >-1.15$). These differences are probably 
due to the population of blue $K$-faint galaxies, that are
missed in the $K$-sample, as discussed in Section \ref{sec:mass}. 
These galaxies have, indeed, median colours 
in the $I$-selected sample that are bluer 
than in the $K$-selected sample ($I-K \simeq 0.45$ compared to $\simeq0.89$).
A similar behaviour has been noted in the local MF derived using
an optically selected sample ($g$ band) compared to the local MF from the 
near-IR (2MASS) sample (Bell et al. 2003).
On the contrary, at even lower masses ($<8.5$ dex) at $z<0.4$ the $K$-selected MF is slightly
steeper than the $I$-selected one, but no significant differences in the colour of the 
two populations is found.

\subsection {Comparison with Previous Surveys}\label{sec:MFcomp}

In general, previous efforts to derive MF have relied on smaller or more 
limited samples, or often based mainly on photometric redshifts (Drory et al. 2004, 2005).
We have compared our MF determination with literature results based
on different surveys (K20, COMBO17, MUSIC, DEEP2, FDF+CDFS), 
rescaled to Chabrier IMF (see Figure \ref{fig:MFlit}).
Our MFs rely on a higher statistics at intermediate to high-mass ranges,
and therefore present lower statistical errors. 
At $z<0.2$ we sample unprecedented mass ranges, more than one order of magnitude lower than 
previous surveys, while at $z>0.4$ the FDF and MUSIC surveys reach lower mass limits even if on 
significantly smaller area.
Our MFs are in fairly good agreement with previous studies
over the whole mass range up to $z\sim 1.2$. However, some differences exist,
in particular at the massive end, which is more sensitive to the different selections,  methods,
statistics and to cosmic variance due to large scale structures: 
for example, in the MUSIC-GOODS survey there are two significant overdensities at $z\sim 0.7$.
The MFs from COMBO17 (Borch et al. 2006) and also from DEEP2 (Bundy et al. \cite{bun06})
are systematically higher than previous surveys 
at the massive-end, in particular in the range $0.7<z<1.2$.
The MFs in the FDF+CDFS are instead systematically lower than ours
at the massive-end and higher than our extrapolation to masses lower than
our completeness limit. 
At $z>1.2$ our MF is systematically higher than previous studies. 
Given the area sampled (more than a factor 4 wider compared to FDF+CDFS and to MUSIC) 
and the consistency at these redshifts of our MF in the 2 $K$-selected separated 
areas (see Figure \ref{fig:MF_shallow_deepK}), we are confident
in our results. 
However at high-$z$
the uncertainties on the stellar masses estimate increase (up to $0.16$ dex including also 
the photometric redshift errors) and could produce a partially spurios excess in the number densities of
galaxies, in particular in the massive tail of the MF. This effect is discussed
in Kitzbichler \& White (2007) which take into account 
in the hierarchical formation Millenium simulation the effect of the dispersion in the mass determination
(0.25 dex, i.e. 78\%). 
We have performed a similar analysis, taking into account the uncertainty on the mass, due to the fitting
technique ($\sim30\%$) and to the uncertainty of the photometric redshifts
(both its dependence on redshift and magnitudes as described in previous sections, i.e. up to
$\sigma_z\simeq0.2$ at $z=2$ and $K>21.5$). We found that the effect on the MF is always small, and 
only the very massive tail ($M>2\times10^{11} M_\odot$) is systematically overestimated (up to 0.2 dex).
This effect can not completely explain the excess found compared to previous surveys, 
which are affected in a similar way by the same bias.

\subsection{The Evolution of the Galaxy Stellar Mass Function}
%\label{sec:evol}

The VVDS allows us to follow the evolution of the MF within a single sample 
over a wide redshift range.
Difficulties in the interpretation of the evolution are, indeed, due to 
the comparison with the local MFs, which have been determined with different 
methods and sample selection.
For example, no local MF has been derived using complex SFHs for mass 
determinations.
In our analysis we use, as reference, the local MF by
Bell et al. (2003) and Cole et al. (2001) rescaled to Chabrier IMF.
In particular, the Cole et al. (2001) local mass function, derived with
smooth SFHs but with formation redshift fixed at $z=20$, has been rescaled to
smooth SFHs method with free formation redshift by Fontana et al. (2004). 

The first important result is that, thanks to our very deep samples, 
both $I$- and $K$-selected,
the low-mass end of the MF is even better determined than in the local sample 
up to $z<0.4$, probing for the first time masses down to about $3 \times 10^7 M_\odot$.
The low mass-end is rather steep ($-1.38 <\alpha<-1.26$), and could even be described by a double 
Schechter function, and is steeper than the local estimates
($\alpha=-1.18\pm0.03$ Cole et al. 2001, 
$\alpha=-1.10\pm0.02$ Bell et al. 2003),
possibly due to the fact they are not probing masses smaller than $10^9$ M$_\odot$
(more than one order of magnitude more massive than in our sample).
As evident from Figure \ref{fig:MF}, we find a substantial population of 
low-mass ($<10^{9}\, M_\odot$) galaxies at
low redshifts ($z<0.4$).  This population is composed by faint blue galaxies
with similar properties in the 2 samples ($I$- and $K$-selected): 
$I, K\simeq 22-23$,
$M_I, M_K \simeq -16, -17$  with median $M_I - M_K \simeq 0.3$,
and median $z\simeq 0.1-0.2$.
This is a very strong result from our
survey which can rely on a wider area and a deeper sample than previous surveys at low redshifts.
At $z>0.4$ the low-mass slope is on the contrary always consistent with the local values.
Even if we are not probing masses smaller than $10^8 M_\odot$ at $z>0.4$, we
found that the MF 
remains quite flat ($-1.23<\alpha<-1.04$) at all redshifts, 
similar to that of Fontana et al. (2006) which probe lower masses 
(see figure \ref{fig:MFlit}).

From a visual inspection of Figure \ref{fig:MF},
we see that up to $z\sim0.9$ there is only a weak evolution of the 
MF, as suggested by previous results (Fontana et al. 2004, 2006, Drory et al. 2005),
while at higher redshifts there appears to begin a decrease in the
normalization of the MF, even if a massive tail remains present up to $z=2.5$.
At intermediate masses ($9.5 <\log M < 10.5$), our VVDS MF is very well defined 
and shows a clear evolution, i.e. the number density decreases with increasing 
redshifts compared to both the first VVDS redshift bin and the local MF.  
This evolution is quite mild up to $z\simeq 0.9$, while it becomes
faster at higher $z$.
At larger masses the high mass end of the MF ($>10^{11}\, M_\odot$) 
shows a small evolution up to $z\simeq 2.5$.
However, its evolution is extremely dependent on
the assumed local MF and on the uncertainties in the mass determination,
which produce a larger dispersion between the different methods and samples
compared to the intermediate-mass range.

%-------------------------------------------------------------------------------
\begin{figure}
\centering
\includegraphics[angle=0,width=\hsize]{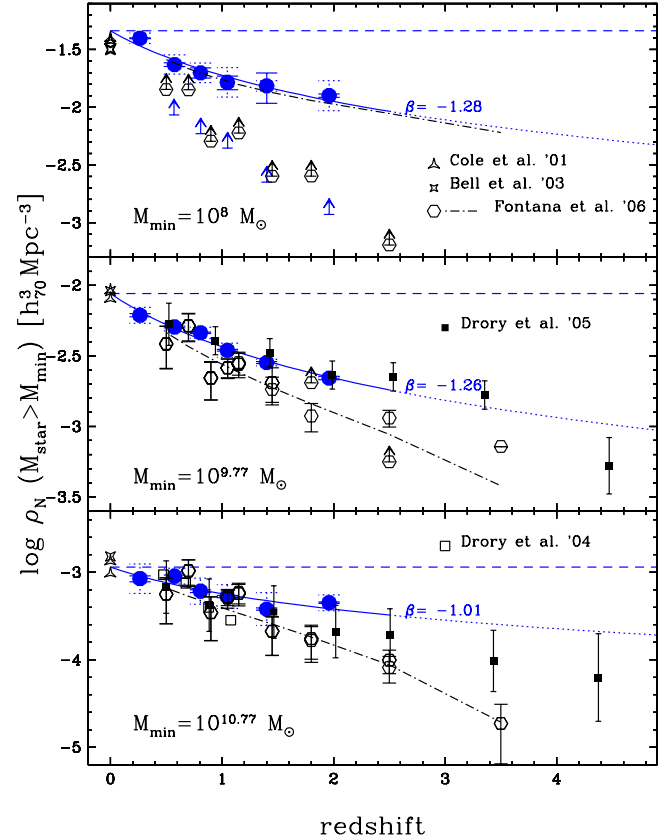}
\caption{Cosmological evolution of the galaxy number density as a function of 
redshift, as observed from the VVDS in various mass ranges
($>10^{8} M_\odot$, $>10^{9.77} M_\odot$ and $>10^{10.77} M_\odot$ 
from top to bottom). 
Observed data from $V_{\rm max}$ (shown as lower limit in the top panel) 
have been corrected,
when necessary, for incompleteness integrating the mass function 
using the best fit Schechter parameters.
VVDS data (big filled circles), averaged
over the $I$- and the $K$-selected samples and the 2 methods to derive the mass,
are plotted along with their statistical errors (solid error bars)
and the scatter between  the 2 different samples and methods (dotted error bars).
The solid lines show the best-fit power laws $\propto (1+z)^{\beta}$, 
while the dashed lines correspond to the no-evolution solution normalized at $z=0$.
Results from previous surveys (small points and dot-dashed lines)
are also shown.
}
\label{fig:NdVz}
\end{figure}
%-------------------------------------------------------------------------------

In order to quantify the MF evolution,  and its mass dependency
independently from the local MF, 
in the next section we derive number densities for different mass limits.

\subsection {Galaxy Number Density}

Here we derive the number density of galaxies as a function of redshift, 
using different lower limits in mass (${\cal M}_{\rm min}$).
We have estimated the number density from the observed
data (from $V_{\rm max}$), as well as
from the incompleteness-corrected MFs, i.e. integrating the
best-fit Schechter functions over the considered mass range.
The corrections due to faint galaxies dominate for ${\cal M}_{\rm min}=10^8 M_\odot$,
while they are negligible for the other mass limits considered.
A formal uncertainty in this procedure was estimated by considering
the $V_{\rm max}$ statistical errors and
the range of acceptable Schechter parameters values.
In Figure \ref{fig:NdVz} we plot our VVDS determinations, averaged over the
$I$- and $K$-selected samples and the 2 methods for mass determination
(listed in Table \ref{tab:ND.MD}),
along with their statistical errors (always less than 10\%) 
and the scatter between the 2 different samples and methods (ranging between 10 and 45\% and due mainly
to the different methods rather than to the different samples).
With the two methods we
find similar trends with redshift of the number densities of galaxies, but
with a systematic shift which is significant only for the highest mass limit 
(for complex SFHs the galaxy number densities are $\sim 50$\% higher than for smooth SFHs).
The effect of photometric redshift and mass uncertainty on the number densities is always small ($<15\%$) 
for the mass range shown in Figure \ref{fig:NdVz}, 
except for the very massive galaxies ($>2\times10^{11}$ $M_\odot$, not shown in the figure because of the 
small number of galaxies in this mass range) where the intrinsic values could be up to a factor 
$\sim 2$ lower (see discussion in 
section \ref{sec:MFcomp}).
The decrease in number density with redshift for all the adopted mass limits
is evident.  

%-------------------------------------------------------------------------------
\begin{table}
\caption[]{Number Density and Stellar Mass Density}
\begin{flushleft}
\begin{tabular}{p{0.4cm} p{0.4cm} p{0.9cm} p{0.7cm} p{0.6cm} p{1.1cm} p{0.7cm} p{0.6cm}} \hline
\hline \\
     $z_{inf}$     & $z_{sup}$      & Log($\rho_N$) & Scatter  & Error    & Log($\rho_{\rm stars}$) & Scatter  &  Error \\ 
\hline 
\hline 
\end{tabular} \\
\begin{tabular}{r}
      log(${\cal M}_{\rm stars}$)$>8$   \\ 
\end{tabular}\\
%\begin{tabular}{p{0.6cm} p{0.6cm} p{1.3cm} p{1cm} p{1.3cm} p{1cm} p{1cm} p{1cm}} \hline
\begin{tabular}{p{0.4cm} p{0.5cm} p{0.8cm} p{0.7cm} p{0.8cm} p{1.0cm} p{0.6cm} p{0.6cm}} \hline
      0.05   &   0.40  &   -1.40  &    0.04   &   0.01  &   8.45  &    0.09   &   0.01  \\
      0.40   &   0.70  &   -1.63  &    0.08   &   0.02   &   8.34  &    0.08   &   0.02  \\
      0.70   &   0.90  &   -1.70  &    0.08   &   0.04   &   8.22  &    0.11   &   0.01  \\
      0.90   &   1.20  &   -1.78  &    0.13   &   0.05   &   8.14  &    0.12   &   0.01  \\
      1.20   &   1.60  &   -1.81  &    0.05   &   0.11   &   8.04  &    0.14   &   0.02  \\
      1.60   &   2.50  &   -1.90  &    0.13   &   0.01   &   8.05  &    0.11   &   0.01  \\
\hline 
\end{tabular} \\
\begin{tabular}{r}
      log(${\cal M}_{\rm stars}$)$>9.77$   \\ 
\end{tabular}\\
%\begin{tabular}{p{0.7cm} p{0.7cm} p{1.3cm} p{1cm} p{1cm} } \hline
\begin{tabular}{p{0.4cm} p{0.5cm} p{0.8cm} p{0.7cm} p{0.9cm} } \hline
%  0.05   &   0.40  & ~~-2.20~~  &  ~0.06~ &  ~~0.01~~  \\
  0.05   &   0.40  & -2.20  &  0.06 &  0.01  \\
      0.40   &   0.70  & -2.29  &   0.03  &   0.01  \\
      0.70   &   0.90  & -2.33  &   0.04  &   0.01  \\
      0.90   &   1.20  & -2.45  &   0.05  &   0.01  \\
      1.20   &   1.60  & -2.54  &   0.02  &   0.01  \\
      1.60   &   2.50  & -2.65  &   0.01  &   0.01   \\
\hline 
\end{tabular}\\
\begin{tabular}{c} 
      log(${\cal M}_{\rm stars}$)$>10.77$   \\ 
\end{tabular}\\
%\begin{tabular}{c c c c c c c c} \hline
\begin{tabular}{p{0.4cm} p{0.5cm} p{0.8cm} p{0.7cm} p{0.9cm} p{1.0cm} p{0.6cm} p{0.7cm}} \hline
% VVDS Stellar Mass Density for logM>10.77 (Pozzetti et al. 2007, astro-ph/0704.1600)
%
%     0.05   &   0.40   &  ~~-3.07~~ &  ~0.18~ &    ~~0.05~~   &   ~~8.00~~   &   ~~~0.22~~~   &   ~0.04~  \\
      0.05   &   0.40   &  -3.07 &  0.18 &    0.05   &   8.00   &   0.22   &   0.04  \\
      0.40   &   0.70   &  -3.04   &   0.10  &    0.02   &   8.04   &   0.14   &   0.02  \\
      0.70   &   0.90   &  -3.22   &   0.16  &    0.02   &   7.80   &   0.19   &   0.02  \\
      0.90   &   1.20   &  -3.28   &   0.15  &    0.02   &   7.76   &   0.18   &   0.02  \\
      1.20   &   1.60   &  -3.42   &   0.23  &    0.02   &   7.59   &   0.28   &   0.02  \\
      1.60   &   2.50   &  -3.35   &   0.11  &    0.01   &   7.70   &   0.11   &   0.01  \\
%
% zmean == mean redshift of the bin (z1<z<z2)
% MD == Stellar Mass density (mean values from 2 sample: I- and K-selected sample 
%                             and 2 methods: smooth SFhs and complex SFHs)
% range+ == upper range of log(MD) covered by the different samples and methods. 
% range- == lower range of log(MD) covered by the different samples and methods. 
% error_MD == statistical error on MD from different samples and methods
% z1 == lower bound of redshift range
% z2 == upper bound of redshift range
% scatter = scatter in log(MD) between the different samples and methods.
%
% H0=70 Omega_matter=0.3, Omega_lambda=0.7, IMF=Chabrier
%
\hline
\end{tabular}
\end{flushleft}
\label{tab:ND.MD}
\end{table}
%-------------------------------------------------------------------------------

We have compared  VVDS results
to previous surveys and  with different local determinations.  
For the total number density ($10^8 < {\cal M}_{\rm stars} < 10^{13} M_\odot$)
VVDS data are very well consistent with the evolutionary STY fit determined by 
Fontana et al. (2006) in the GOODS-MUSIC survey.
At intermediate masses ($>10^{9.77} M_\odot$, corresponding 
to $10^{10} M_\odot$ for Salpeter IMF)
our VVDS data
have a better determination and smaller uncertainties than previous ones and are
consistent with most of them at $z<1.2$ and in the upper envelope at higher $z$.
For the high mass range ($>10^{10.77} M_\odot$, corresponding 
to $10^{11} M_\odot$ for Salpeter IMF) we are quite consistent with previous results, 
and even if our VVDS have lower errors than previous ones, 
the dispersion within the various VVDS measurements reflect the uncertainties
for massive galaxies.
%-------------------------------------------------------------------------------
\begin{figure}
\centering
\includegraphics[angle=0,width=\hsize]{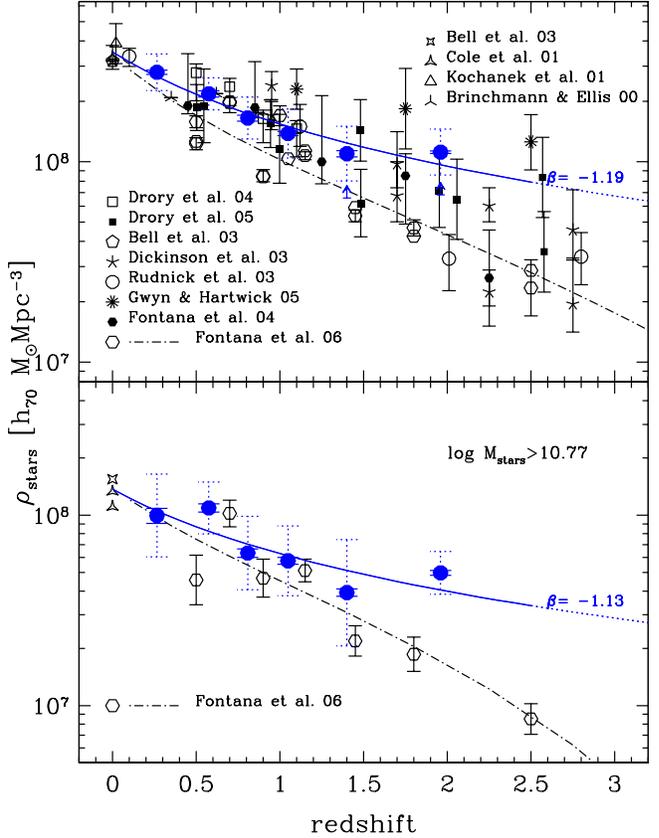}
\caption{Cosmological evolution of the stellar mass density as a function of 
redshift as observed from the VVDS for 2 mass ranges: integrated over
the whole range $10^{8} M_\odot \leq {\cal M}_{\rm stars} \leq 10^{13} M_\odot$ (upper panels) 
and for massive galaxies ($>10^{10.77} M_\odot$) (lower panels).
Symbols and lines as in Figure \ref{fig:NdVz}.
}
\label{fig:MD}
\end{figure}
%-------------------------------------------------------------------------------

If we represent the average number density evolution by a power law 
$\rho_N\propto (1+z)^{\beta}$, we find 
that $\beta({\cal M}_{\rm stars}>10^{8})=-1.28\pm 0.15$, $\beta({\cal M}_{\rm stars}>10^{9.77})=-1.26\pm0.10$,
and $\beta({\cal M}_{\rm stars}>10^{10.77})=-1.01\pm0.05$ 
(the errors on $\beta$ represent the uncertainties due to the 2 different methods).
We find on average a similar evolution for the 2 methods analyzed
and a slightly milder evolution with increasing mass limit (`downsizing' in mass assembly).
The average evolution from $z=0$ to $z=1$ is a factor $2.4\pm0.3$ and $2.0\pm0.1$ 
from low to high-mass galaxies, respectively,
and increases to a factor $4.0\pm0.9$ and $3.0\pm0.2$, respectively, at $z=2$.
We note moreover that for the highest mass limit (${\cal M}_{\rm stars}>10^{10.77}$)
at low redshift
($z<0.7$) the number density observed is consistent with no-evolution 
(fixing the value to $z=0$ we found an evolution $<30\%$),
excluded instead for the intermediate and low-mass limit.
At high-$z$ ($z>1.5$) for intermediate and high-mass range 
we note a flattening in the number density from the VVDS data, 
similar to Drory et al. (2005), but higher than Fontana et al. 
(2006). This flattening is
due to the high mass tail observed in the range $1.6<z<2.5$. 
This population shows extremely red colours ($M_I-M_K\simeq 0.8$)
and could be related to the appearance of a population of massive star forming dusty 
galaxies, observed in previous surveys (Fontana et al. 2004, Daddi et al. 2004b). 
The small excess induced by uncertainties on the mass and photometric redshifts 
(see discussion in section \ref{sec:MFcomp})
can not completely explain the difference with Fontana et al.
(2006) survey, which is affected by the same bias in a similar way.

The mass dependent evolution (``mass downsizing") is very 
debated and the results from different surveys are still controversial. Deep
surveys, such as the FDF \& CDFS (analyzed by Drory et
al. 2005) find an evolution consistent with ours
(a decrease of about a factor 2.5 -- 4 at $z=1-2$
in the number density of galaxies $>10^{10.77}\, M_\odot$).
Fontana et al. (2006) suggest a similar mild evolution up to $z=1$,
for massive galaxies ($>10^{10.77}\, M_\odot$) 
and a stronger evolution at $z>1.5$, reaching a factor about 10 at $z=3$.
Similarly,  Cimatti et al. (2006) show that the 
number density of luminous (massive, ${\cal M}_{\rm stars}>10^{11}\, M_\odot$) early-type galaxies is nearly constant 
up to $z\sim0.8$, while Bundy et al. (2006) find a slight decrease, consistent
with no evolution, only for even more massive system ($>3\times10^{11}\, M_\odot$)
and a more significant decline for ${\cal M}_{\rm stars}<3\times10^{11}\, M_\odot$.
Vice versa, data from the MUNICS survey (Drory et al. 2004)
show a faster evolution of massive galaxies, even faster than for the less 
massive systems (see also Figure 4 in Drory et al. 2005).

To summarize, our accurate results show that the MF evolves mildly
up to $z\simeq1$ (about a factor 2.5 in the total number density) and that a high-mass tail 
is still present up to $z=2.5$.
Moreover, we find that massive systems show an evolution that is 
on average milder ($<50$\% at $z<1$) than 
intermediate and low-mass galaxies and 
consistent with a mild/negligible evolution ($<30$\%) up to $z\sim0.8$.
Conversely, a no-evolution scenario in the same redshift range 
is definitely excluded for intermediate- and low-mass galaxies.

This behaviour suggests that 
the assembly of the stellar mass in 
objects with mass smaller than the local ${\cal M}_{\rm stars}^*$ was quite
significant between $z=2$ and $z=0$. Qualitatively, this behaviour
is expected for galaxies with SFHs prolonged over cosmic time,
which therefore continue to grow in terms of stellar mass after $z\sim1$.
Conversely, our results further strengthen the fact that
the number density of massive galaxies is roughly constant up to 
$z\simeq 0.8$, consistently with a SFH peaked at higher
redshifts, with the conversion of most of their gas into stars
happening at $z>1.5-2$,
ruling out
the `dry mergers' as the major mechanism of their assembly history, below $z<1$.

\section{Mass Density}\label{sec:MD}

Various attempts to reconstruct the cosmic evolution of the 
stellar mass density have been previously made, mainly using NIR-selected samples
(Dickinson et al. 2003, Fontana et al. 2004, Drory et al. 2005).
Our survey offers
the possibility to investigate it using the MF derived from 
two different optical- and NIR-selected samples, taking advantage of our depth,
and relatively wide area covered. Furthermore, the different methods analyzed here to derive the 
stellar mass content give us a direct measure of the uncertainties involved. 

We have estimated the stellar mass density from the
observed data, as well as from the incompleteness-corrected MFs.
Up to $z<1$ the corrections due to faint galaxies 
are relatively small. 
A formal uncertainty in this procedure was estimated by considering
the $V_{\rm max}$ statistical errors and
the range of acceptable Schechter parameters values.

Figure \ref{fig:MD} shows our results (averaged over the $I$- and $K$-selected samples
and the 2 methods with a typical scatter of about 30-50\% and statistical errors always less than 5\%
see Table \ref{tab:ND.MD}),
for the total mass density and for the density in massive 
galaxies ($>10^{10.77}$),
along with their representative power laws ($\rho_{\rm stars}\propto (1+z)^{\beta}$),
and compared to literature data (see references in the figure).
For the total mass density, even if the results from our survey cover a range of values with
some significant differences between the two different methods (up to $\sim 40$\%), 
the general behaviour and evolutionary trend is well defined by
$\beta=-1.19\pm0.05$.  
We find
that the evolution of the stellar mass density
is relatively slow with redshift, with a decrease of a factor
$2.3\pm0.1$ up to $z\simeq 1$,
up to a factor $4.5\pm0.3$ at $z=2.5$.
The agreement of average total mass density with previous surveys is 
reasonably good, 
and the range covered by VVDS data reflect the different selection 
techniques and methods used in different surveys.
The average total mass density evolution
is milder than in the MUSIC sample (Fontana et
al. 2006) already at $z>0.5$. Our evolutionary trend is consistent with the upper envelope of 
previous surveys, even if our highest-redshift value is uncertain because the low-mass slope 
is poorly constrained.
For comparison the analysis of VVDS data using a IRAC-selected sample (see Arnouts et al. 2007)
finds similar values for the mass density, except that the highest redshift point is lower than ours.
Given the present uncertainties on the low-mass slope of the GSMF, the total mass density
at $z\simeq2$ remains poorly constrained.  

The mass density of high-mass objects ($>10^{11} M_\odot$ with Salpeter IMF) varies by a factor
up to $1.8$ within the 2 methods adopted, but the evolutionary trend is similar
($\beta=-1.13\pm 0.01$) and 
consistent with a decrease of about a factor $2.18\pm0.02$ to $z=1$ and $3.44\pm 0.04$ to $z=2.0$.
Moreover at low redshift ($z<0.7$) the VVDS observed data are consistent with a mild/negligible evolution
($<30$\%), as indicated by the number density of massive galaxies (see previous Section).
Our data are roughly 
consistent with Fontana et al. (2006) up to $z=1.5$ 
(even if the slope of the evolutionary trend is shallower),
while at $z>1.5$ the VVDS mass-density of massive galaxies 
is significantly higher than that in
Fontana  et al. (2006), reflecting the excess in MF at high-$z$ noted in 
the VVDS MF compared to previous ones (see Section \ref{sec:MFcomp}).
This results, therefore, in a flatter evolutionary trend over the total redshift range.
Given the wider area and completeness for high-mass objects, our samples
guarantee a higher statistical accuracy and confidence level than before.
However some caveat remains due to the effect of photometric redshift and mass uncertainty on 
the mass densities, which is anyhow always small ($<15\%$) 
except for the massive galaxies ($>6\times 10^{10}$ $M_\odot$) where the intrinsic values could 
be up to 20-30\% lower than our estimates (see discussion in Section \ref{sec:MFcomp}).

\section{Summary and Discussion}\label{sec:concl}

We have investigated the evolution of the Galaxy Stellar Mass Function
up to to $z=2.5$ using the VVDS survey covered by deep VIMOS spectroscopy
($17.5<I<24$) and multiband photometry (from $U$ to $K$-band). 
For our analysis we have used two different samples:
(1) the optical ($I$-selected, $17.5<I<24$) main spectroscopic sample,
based on about 6500 secure redshifts over about 1750 arcmin$^2$,
and (2) a near-IR sample ($K$-selected, $K<22.84$ \& $K<22.34$), in a sub-area 
of about 610 arcmin$^2$ and based on about 10200
galaxies with accurate photometric and spectroscopic redshifts. 
For the first time we have probed masses down to 
a very low limit, in particular at low-$z$ (down to $\sim 3\times10^7 M_\odot$ at $z\sim 0.2$),
while the relatively wide area has allowed us to determine the MF with much higher statistical accuracy 
than previous samples.

In order to better understand uncertainties we have applied and compared
two methods to estimate the stellar mass content in galaxies from
multiband SED fitting. The 2 methods differ in the explored parameter space 
(metallicity, dust law and content) 
and are based on different assumptions on previous star formation history.
The main results from the stellar mass estimate can be summarized 
as follows:

\begin{itemize}

\item The agreement between the 2 methods 
is fairly good even if 
masses estimated with `complex SFHs' are 
systematically higher than `smooth SFHs' masses.
For the $K$-selected sample
the mean difference is
$\langle d\log {\cal M}_{\rm stars} \rangle\simeq 0.12$ dex, and the dispersion is $\sigma=0.13$. 
The differences are
mainly due to the secondary burst component (complex SFHs) 
compared to smooth SFHs. 

\item We found that mass estimates using only optical
bands are in rather good agreement with those using also NIR bands up
to $z\sim1.2$.
We have used this information to statistically correct masses for
objects without near-IR photometry.
At higher redshifts the shift and dispersion dramatically increase 
and the mass estimates become unreliable if near-IR photometry is not available.

\end {itemize}

We have, thus, derived the MF using the VVDS $I$-selected sample and 
extended it up to $z=2.5$ thanks to the $K$-selected sample.
From a detailed analysis of the MF, galaxy number density  and mass density,
in different mass ranges, through cosmic
time, we found evidences for:

\begin{itemize}

\item a substantial population of low-mass galaxies ($<10^9 M_\odot$)  at $z\simeq0.2$
composed by faint ($I,K \simeq 22,23$) blue galaxies with median $M_I-M_K \simeq 0.3$, 
and absolute magnitudes $M_I, M_K \simeq -16, -17$;

\item a slow evolution of the stellar mass function with redshift up to $z\sim0.9$
and a faster evolution at higher-$z$, in particular for less massive systems.
A massive population is present up to $z=2.5$ and have extremely red colours ($M_I-M_K\simeq 0.7-0.8$).

\item at $z>0.4$ the low-mass slope of the GSMF does not evolve significantly and 
remains quite flat ($-1.23<\alpha<-1.04$).

\item the number density shows, on average, a mild differential evolution with mass, 
which is slower with increasing mass limit. 
Such evolution can be described by a power law
$\propto (1+z)^{\beta(>M)}$. Within the VVDS redshift range we found that 
$\beta(>10^{8}\, M_\odot)=-1.28\pm 0.15$, $\beta(>10^{9.77}\, M_\odot)=-1.26\pm 0.10$ and 
$\beta(>10^{10.77}\, M_\odot)=-1.01\pm 0.05$.
For massive galaxies at low redshift ($z<0.7$) the evolution is consistent with 
mild/negligible-evolution ($<30$\%), which is excluded for low-mass systems.

\item the evolution of the stellar mass density
is relatively slow with redshift, with a decrease of about
a factor $2.3\pm0.1$ to  $z\simeq1$, while
at $z\simeq2.5$ the decrease amounts to a factor up to $4.5\pm0.3$,
milder than in previous surveys.
For massive galaxies the evolution at low redshift ($z<0.7$) is consistent with a 
mild/negligible evolution($<30\%$), and shows a flattening compared to previous results at $z>1.5$
due to a population with extremely red colours.

\end{itemize}

Our results provide new clues on the controversial question 
of when galaxy formed and assembled their stellar mass.
Most of the massive galaxies seem to be in place up to $z=1$ 
and have, therefore, formed their stellar mass at high redshift 
($z>1$), 
rather than assembled it mainly through continuous galaxy merging of small galaxies
at $z<1$.
On the contrary, less massive systems have assembled their mass
(through merging or prolonged star formation history) later in cosmic time.
In agreement with our results, a substantial population of high-$z$ 
($z\sim2-3$) dusty and massive objects have been discovered in near-IR surveys
(Daddi et al. 2004b) and detected by Spitzer in the far-IR (Daddi et al. 2005, Caputi et al. 2006b). 
This population
could be related to the initial phase of massive galaxy formation
during their strong star forming and dusty phase. 

Finally, our results are not completely accounted for by most of 
theoretical models of galaxy formation (see Fontana et al. 2004,  2006
and Caputi et al. 2006a for a detailed comparison with models). For instance,
models by De Lucia et al. (2006) predict that the most massive galaxies generally
form their stars earlier, but assemble them later, mainly at $z<1$ via merging,
than the less massive galaxies (i.e. 'downsizing in star formation but 'upsizing'
in mass assembly, see Renzini 2007 for a recent discussion).
Furthermore, the stronger decrease with redshift of the low-mass population, with a low-mass end of the GSMF 
which remains substantially flat up to high redshift,
is not reproduced by most of the theoretical galaxy assembly models, 
which tend, indeed,  to overpredict the low-mass end of the MF (see Fontana et al. 2006).

Understanding the mass assembly of less massive objects
and disentangling merging processes from prolonged star formation history
is more complicated.
In this respect for a better comprehension of galaxy formation the 
VVDS will allow us to further investigate the evolution of the stellar mass 
function up to high-$z$ also for different galaxy types 
(spectral and morphological) and in different environments.
For example Arnouts et al. (2007) 
study the mass density evolution of different galaxy population.
Further analysis of galaxy mass dependent evolution, using 
stellar population properties, as well as observed spectral features, 
will be presented in forthcoming papers 
(Lamareille et al., in preparation, Vergani et al. 2007). 
Furthermore, 
it will be possible to push the study of the galaxy stellar mass 
function at higher redshifts using SPITZER mid-IR observations.
While most of present studies (Dickinson et al. 2003, Drory et al. 2005) 
do not use rest-frame near-IR photometry to estimate stellar masses,
our VVDS-SWIRE collaboration will allow to combine the deep VVDS spectroscopic 
sample with SPITZER-IRAC photometry.

\begin{acknowledgements}
This research has been developed within the framework of the VVDS
consortium.\\
This work has been partially supported by the
CNRS-INSU and its Programme National de Cosmologie (France),
and by Italian Ministry (MIUR) grants
COFIN2000 (MM02037133) and COFIN2003 (num.2003020150).\\
The VLT-VIMOS observations have been carried out on guaranteed
time (GTO) allocated by the European Southern Observatory (ESO)
to the VIRMOS consortium, under a contractual agreement between the
Centre National de la Recherche Scientifique of France, heading
a consortium of French and Italian institutes, and ESO,
to design, manufacture and test the VIMOS instrument.
We are in debt with E. Bell, S. Salimbeni, and E. Fontana for providing 
the data from their survey in electronic format, and to C. Maraston 
for her galaxy evolution models in BC format.

\end{acknowledgements}

\end{document}